\documentclass[journal]{IEEEtran}
\ifCLASSINFOpdf
\else
\fi
\usepackage{multirow}
\usepackage{url}
\usepackage{enumerate}
\usepackage{graphicx}
\usepackage{amsmath,amssymb}
\usepackage{bm}
\usepackage[algo2e]{algorithm2e}
\usepackage{algorithm,algpseudocode}
\begin{document}
%
\title{\huge Revenue and Energy Efficiency-Driven Delay Constrained Computing Task Offloading and Resource Allocation in a Vehicular Edge Computing Network: A Deep Reinforcement Learning Approach}
%
%
%
\author{Xinyu Huang, Lijun He$^{*}$, Xing Chen, Liejun Wang, Fan Li
	
	\thanks{X. Huang, L. He, X. Chen and F. Li are with the School of Information and Communications Engineering, Xi'an Jiaotong University, Xi'an 710049, China (e-mail: \{xinyu\_huang, xingchen,\}@stu.xjtu.edu.cn; \{lijunhe, lifan\}@mail.xjtu.edu.cn). \textit{(Corresponding author: Lijun He, email: lijunhe@mail.xjtu.edu.cn)}
		
	L. Wang is with the College of Information Science and Engineering, Xinjiang University, Urumqi 830046, China (email: wljxju@xju.edu.cn).
	}}

\maketitle

\begin{abstract}
For in-vehicle application, task type and vehicle state information, i.e., vehicle speed, bear a significant impact on the task delay requirement. However, the joint impact of task type and vehicle speed on the task delay constraint has not been studied, and this lack of study may cause a mismatch between the requirement of the task delay and allocated computation and wireless resources. In this paper, we propose a joint task type and vehicle speed-aware task offloading and resource allocation strategy to decrease the vehicle's energy cost for executing tasks and increase the revenue of the vehicle for processing tasks within the delay constraint. First, we establish the joint task type and vehicle speed-aware delay constraint model. Then, the delay, energy cost and revenue for task execution in the vehicular edge computing (VEC) server, local terminal and terminals of other vehicles are calculated. Based on the energy cost and revenue from task execution, the utility function of the vehicle is acquired. Next, we formulate a joint optimization of task offloading and resource allocation to maximize the utility level of the vehicles subject to the constraints of task delay, computation resources and wireless resources. To obtain a near-optimal solution of the formulated problem, a joint offloading and resource allocation based on the multi-agent deep deterministic policy gradient (JORA-MADDPG) algorithm is proposed to maximize the utility level of vehicles. Simulation results show that our algorithm can achieve superior performance in task completion delay, vehicles' energy cost and processing revenue.
\end{abstract}


%
\IEEEpeerreviewmaketitle

\section{Introduction}
%
%
%
%
With the rapid development of the Internet of Things, smart in-vehicle applications (i.e., autonomous driving, image-assisted navigation and multimedia entertainment) have been widely applied to smart vehicles \cite{Shi}, \cite{Zhang}. These applications can provide a more comfortable and safer environment for drivers and passengers. Nevertheless, these in-vehicle applications will consume much computation resources and require low execution latency, which is difficult to satisfy by the vehicle itself due to the limited computing capability of the vehicle. Luckily, VEC servers with powerful computing capacity are deployed densely on the roadside and attached in the roadside unit (RSU) \cite{Pu}, which provides a tremendous potential for satisfactory performance of in-vehicle applications such as: 1) Low latency: The VEC server is closer to the vehicle terminals and possesses powerful computing capacity, which can help vehicle terminals process the delay-intensive tasks quickly and decrease the traffic burden on the backhaul network. 2) Energy efficiency: Energy consumption is a key factor for vehicles equipped with resource-hungry applications. By offloading the computation-intensive tasks to the VEC server, the energy consumption of the vehicles can be maintained at a low level. 3) Context awareness: The VEC server can obtain the state information of the vehicles, i.e., vehicle position, vehicle speed, task queue, channel state and remaining computation and wireless resources, etc., in real time. Therefore, VEC-based optimization schemes can more comprehensively optimize task offloading and resource allocation from the system perspective. Currently, VEC-based optimization schemes have focused mainly on how to offload computing tasks to a VEC server to achieve the goal of low latency and energy efficiency. However, how the variation of the vehicles' information such as vehicle speed and task type affects the requirements of in-vehicle applications has not been extensively explored. Even for the same in-vehicle application, different vehicle speeds in different scenarios require different delay constraints. Therefore, it is worth considering the information for these vehicles during offloading and resource allocation in a VEC network to satisfy different requirements of in-vehicle applications. 

Many researchers have made extensive innovations on the architecture and function modules of the VEC and have discussed the potential challenge and prospects of optimization of VEC-based task offloading. The authors in \cite{Raza} proposed a three-layer vehicular edge computing architecture and analyzed the role and mode of operation of each component in detail. This work can be regarded as a guide for future research on optimization of VEC-based task offloading. In another respect, \cite{Feng} established a vehicular edge computing platform of the workflow for vehicles and management for idle computation resources to increase the computing capabilities of vehicles in a decentralized manner. To further improve the efficiency of task offloading, the authors in \cite{Feng2} proposed to utilize the vehicle-to-vehicle (V2V) communication to offload computing tasks to reduce the processing burden on the VEC server and discussed the challenges of offloading decision and resource allocation for resource-hungry and computation-intensive applications. For these in-vehicle applications, the authors in \cite{Dziyauddin} conducted a comprehensive survey on current task offloading methods and presented future research directions of VEC-based task offloading optimization. Generally, the main types of VEC-based offloading optimization can be summarized as the optimization of delay-driven, energy-driven, revenue-driven and multi-objective driven offloading and resource allocation. By optimizing computing task offloading methods, the requirements of different in-vehicle applications can be satisfied to improve the reliability of autonomous driving and the quality of experience (QoE) in VEC network for the user.

Although the above-mentioned work has demonstrated significant efforts in the performance improvement of in-vehicle applications, there are still some problems to be solved: (1) The effect of vehicle speed on task delay has not been studied, resulting in a mismatch between vehicle speed and the allocated computation and wireless resources. (2) The vast fluctuation of the vehicle channel state caused by fast mobility has not been extensively explored, which can cause task offloading to suffer failure when allocated wireless channels cannot guarantee successful data transmission. (3) State information for vehicles is large and complex, which makes it difficult to quickly obtain the optimal offloading and resource allocation strategy for each vehicle. In addition, the optimization problem is generally related to multivariate nonlinear mixed integer programming, presenting a challenge to find the optimal solution. (4) The mmwave-based V2V communication has not been extensively studied in the collaborative offloading mechanism among vehicles, which results in a low transmission capacity when computing tasks are offloaded from one vehicle to another. (5) The relationship among task completion delay, energy cost and revenue for processing tasks has not been jointly explored in the task offloading and resource allocation strategy, which can produce a low efficiency of task offloading in a VEC-based network.

Inspired by the above problems, we propose a revenue and energy efficiency-driven delay constrained computing task offloading and resource allocation strategy based on multi-agent deep reinforcement learning. Our work is novel in the following aspects.

\textit{ 1) Comprehensive VEC-based computing task offloading framework:}

 In contrast to the VEC-based framework of only offloading tasks to the VEC server, we propose to establish a comprehensive framework that includes V2V and vehicle to infrastructure (V2I) offloading patterns, and computation and wireless resource allocation in vehicle side and VEC server. The establishment of this framework can help the VEC server to collect and utilize the state information of the vehicles to make the task of offloading and resource allocation strategy for each vehicle improve the performance of the VEC-based offloading network.	

\textit{ 2) Vehicle speed and task type-aware delay constraint model:}

 Based on the bandwidth and delay requirements, in-vehicle computing tasks are classified into three types: critical application, high-priority application and low-priority application. Different types of tasks and vehicle speeds demand various delay requirements for in-vehicle applications. Therefore, we fully analyze the internal relationship among vehicle speed, task type and delay requirements to propose a vehicle speed and task type-aware delay constraint model, which can make the task of offloading and the resource allocation process more accurate.

\textit{ 3) Refined task offloading for various offloading patterns:}

 Different types of tasks in the task queue can be handled in four ways, which are hold on, offloaded to VEC server, offloaded to other vehicles and local execution. For different offloading patterns, we select the task completion delay, energy cost and revenue to evaluate the refined task-offloading procedure. To be specific, we establish the delay model based on the channel resource and computation resource allocation, the energy cost model based on the transmission and execution cost, and the revenue model based on the collaborative offloading mechanism, respectively. 

\textit{ 4) Joint optimization of task offloading and resource allocation based on multiagent reinforcement learning algorithm:}

 Since the offloading and resource allocation variables consist of high-dimension matrixes, it is difficult for traditional methods to quickly obtain the optimal offloading and resource allocation strategy for each vehicle. To address this issue, we employ the deep reinforcement learning method to solve this problem. Specifically, a joint optimization of offloading and resource allocation problem is formulated by the Markov decision process (MDP), with the objective of maximizing the utility of vehicles subject to delay constraint. Based on the objective function and constraints, the reward function is well designed to accelerate the training process. Eventually, by directly utilizing a pretrained model, the VEC server can quickly determine the near-optimal offloading and resource allocation strategy for each vehicle.

The rest of the paper is organized as follows: Section II analyzes the development of the related work. Section III shows the system framework and provides the joint task type and vehicle speed-aware delay constraint model. Section IV presents the derivation procedure of delay, energy cost and revenue of different offloading patterns. Section V introduces the joint optimization of computing task offloading and resource allocation, which consists of the mathematical model formulation and the problem solution. Section VI provides the numerical results of the proposed algorithms and some of the existing algorithms. Finally, we draw the conclusion of the paper in Section VII.

\section{Related Work}
To enhance the efficiency of computing task offloading and improve the QoE of drivers and passengers in a VEC-based network, researchers have proposed many VEC-based offloading and resource allocation optimization algorithms, which can be divided into four categories: delay-driven optimization algorithms, energy-driven optimization algorithms, revenue-driven optimization algorithms and multi-objective optimization algorithms.

\subsection{Delay-driven optimization algorithms}
Low latency of computing task is essential to guarantee that vehicles can complete corresponding actions to ensure the safety of drivers and passengers. To reduce computing task completion delay, many researchers have proposed delay-driven task offloading optimization methods. The authors in \cite{Sun} and \cite{Sun2} developed a learning-based task offloading framework using the multi-armed bandit (MAB) theory, which enables vehicles to learn the potential task offloading performance of its neighboring vehicles with excessive computation resources to minimize the average offloading delay. However, more task vehicles should be considered rather than the fixed task vehicle in the task offloading scenario. To further improve the offloading cooperation among vehicles, the authors in \cite{Qiao} proposed a collaborative task offloading and output transmission mechanism to guarantee low latency and load-balancing but ignored the delay requirements of different types of tasks. In addition, the authors in \cite{Liu} focused on different road conditions and utilized a delay-based task offloading algorithm to minimize the total network delay. However, the impact of vehicle mobility on the change of channel state should be considered in a more detailed manner in the task-offloading process.

\subsection{Energy-driven optimization algorithms}
Computation-intensive applications usually consume a large amount of the energy of the vehicle, which is a disaster, especially for the electric car. To reduce the energy cost to the vehicle of processing computing tasks, many researchers proposed energy-driven task offloading optimization methods. The authors in \cite{Zhou} proposed a joint task offloading and power control problem, with the explicit consideration of energy consumption and delay models. An alternating direction method of multipliers (ADMM)-based energy-efficient resource allocation algorithm was developed to reduce the energy consumption of the vehicles as soon as possible. However, the cooperative offloading mechanism among vehicles was neglected. Considering the V2V offloading pattern, the authors in \cite{Wang} proposed a semi-Markov decision process (SMDP)-based cooperation strategy, where the bus-based cloudlets acted as computation service providers for in-vehicle applications to minimize the energy consumption of vehicles within task delay constraints. To further reduce energy consumption, the authors in \cite{Mu} utilized dynamic programming to obtain the optimal solution to the assignment and scheduling problem. However, these two methods neglected the condition that the waiting task can still select to hold on until there are sufficient computation and wireless resources to process within the delay constraint.  

\subsection{Revenue-driven optimization algorithms}
The revenue of operator and vehicles by providing a task offloading service is essential to create the optimal offloading schemes for each vehicle. To improve the system revenue, many researchers proposed revenue-driven task offloading optimized methods. The authors in \cite{Zhang2} adopted a contract theoretical approach to design optimal offloading strategies for the VEC service provider, which maximized the revenue of the provider while enhancing the utility of the vehicles. A mobility-aware and a location-based task offloading were introduced in \cite{Yang} to minimize the completed system cost, including the communication and computation costs of the required vehicle while satisfying the task latency. However, these two methods neglected the different delay requirements of various types of in-vehicle applications. Considering the different delay requirements, the authors in \cite{Zhang3} employed the contract-based computation resource allocation algorithm to maximize the benefit to the VEC service provider. To further improve the system revenue, the authors in [18] adopted a Stackelberg game theoretical approach to design an optimal multilevel offloading scheme, which achieved a revenue balance of both the vehicles and the VEC servers. However, these two methods did not apply V2V offloading to the computing task offloading scenario.

\subsection{Multi-objective optimization algorithms}
To improve the overall performance of the VEC network, many researchers comprehensively considered different evaluation indices to formulate the multi-objective task offloading optimized methods. The authors in \cite{Sun3} proposed a novel multi-objective vehicular edge computing task scheduling algorithm by jointly optimizing the allocation of communication and computing resources to minimize total execution time and to maximize total successful tasks. In addition, considering both energy consumption and execution delay, an optimization problem was formulated to maximize the QoE of the vehicles and then solved by the improved double-deep Q networks \cite{Ning}. To further decrease the energy consumption and execution delay, the authors in \cite{Zhan} designed the task offloading algorithm based on the proximal policy optimization (PPO) algorithm to solve the problem of where to schedule and when to schedule, which can achieve a trade-off between task latency and energy consumption. In addition, the authors in \cite{Ke} proposed an adaptive offloading method based on the deep deterministic policy gradient (DDPG) to minimize the total cost of the energy consumption and data transmission delay. However, these three methods neglected the V2V offloading for the computing task. In \cite{Liu2}, the communication utility and computation utility of the VEC network are jointly optimized to maximize the system revenue while guaranteeing the delay of the computing task. However, the impact of vehicle speed on different tasks was neglected, which may cause the high-speed vehicle to be unable to obtain the sufficient computation and wireless resources to complete crucial computing tasks. In addition, the authors in \cite{Zhao} selected task processing delay, cost of computation resource, and the normalization factor to constitute the system utility. The game-theoretical approach was employed to make offloading decisions, and the Lagrange multiplier method was utilized to obtain the resource allocation. However, the V2V offloading and the impact of vehicle speed on the task delay constraint were still not considered in the computing task offloading and resource allocation.

\section{System Model}
This section describes the system model in detail, which includes system framework and joint task type and vehicle speed-aware delay constraint model. The main notations used in this work are listed in Table \ref{tab:not}.

\begin{table}[!h]\scriptsize
	\caption{The list of main notations}\label{tab:not}
	\begin{tabular}{|p{1cm}|p{7cm}|}
		\hline
		\multicolumn{1}{|c|}{\textbf{Notation}}    & \multicolumn{1}{c|}{\textbf{Definition}}                                                                                                                                                                  \\ \hline
		${{\mathcal{I}}_{t,k}}$                    & The computing task of the vehicle $k$ at time $t$ to be   processed                                                                                                                                       \\ \hline
		$\Upsilon ({{v}_{{{\mathcal{I}}_{t,k}}}})$ & The delay threshold of computing task ${{\mathcal{I}}_{t,k}}$                                                                                                                                             \\ \hline
		$\mathbb{Q}$                               & The set of vehicles in the service area of VEC server                                                                                                                                                     \\ \hline
		$K$                                        & The number of vehicles served by VEC server                                                                                                                                                               \\ \hline
		$rv_{k,n}^{{}}$                            & \begin{tabular}[c]{@{}l@{}}The available uplink transmission capacity when uplink channel $n$ is \\ allocated to vehicle $k$ for the communication between vehicle $k$ \\ and VEC server\end{tabular}     \\ \hline
		$\widetilde{rv}_{k,n}$                                          & \begin{tabular}[c]{@{}l@{}}The available downlink transmission capacity when downlink channel \\ $n$ is allocated to vehicle $k$ for the communication between vehicle $k$ \\ and VEC server\end{tabular} \\ \hline
		$rc_{k,n}^{{{k}^{'}}}$                     & \begin{tabular}[c]{@{}l@{}}The available uplink transmission capacity when uplink channel $n$ is \\ allocated to vehicle $k$ for the communication between vehicle $k$ \\ and vehicle $k'$\end{tabular}    \\ \hline
		$\widetilde{rc}_{k,n}^{{{k}^{'}}}$                                          & \begin{tabular}[c]{@{}l@{}}The available downlink transmission capacity when downlink channel \\ $n$ is allocated to vehicle $k$ for the communication between vehicle $k$ \\ and vehicle $k'$\end{tabular} \\ \hline
		$T{{R}_{{{\mathcal{I}}_{t,k}}}}$           & The completion time of task ${{\mathcal{I}}_{t,k}}$   when offloaded to VEC server                                                                                                                        \\ \hline
		$TV_{{{\mathcal{I}}_{t,k}}}^{{{k}^{'}}}$   & The completion time of task when ${{\mathcal{I}}_{t,k}}$   offloaded to vehicle $k’$                                                                                                                        \\ \hline
		$T{{L}_{{{\mathcal{I}}_{t,k}}}}$           & The completion time of task ${{\mathcal{I}}_{t,k}}$   when executed locally                                                                                                                               \\ \hline
		$D({{\mathcal{I}}_{t,k}})$                 & The total delay of computing task ${{\mathcal{I}}_{t,k}}$ from   generation to completion                                                                                                                 \\ \hline
		$E{{R}_{{{\mathcal{I}}_{t,k}}}}$           & The energy consumption of task ${{\mathcal{I}}_{t,k}}$ offloaded   to VEC server                                                                                                                          \\ \hline
		$EV_{{{\mathcal{I}}_{t,k}}}^{{{k}^{'}}}$   & The energy consumption of task ${{\mathcal{I}}_{t,k}}$ offloaded   to vehicle $k’$                                                                                                                          \\ \hline
		$E{{L}_{{{\mathcal{I}}_{t,k}}}}$           & The consumed energy of task ${{\mathcal{I}}_{t,k}}$   when executed locally                                                                                                                               \\ \hline
		${{E}_{t}}$                                & The consumed energy of all vehicles served by VEC   server at time $t$                                                                                                                                      \\ \hline
		$TR_t$                                           & The total actual revenue of all vehicles at time $t$                                                                                                                                                        \\ \hline
		${{U}_{t}}$                                & The utility function of all vehicles at time $t$                                                                                                                                                            \\ \hline
	\end{tabular}
\end{table}

\subsection{System Framework}
\begin{figure}[!h]
	\centering
	\includegraphics[width=8.8cm]{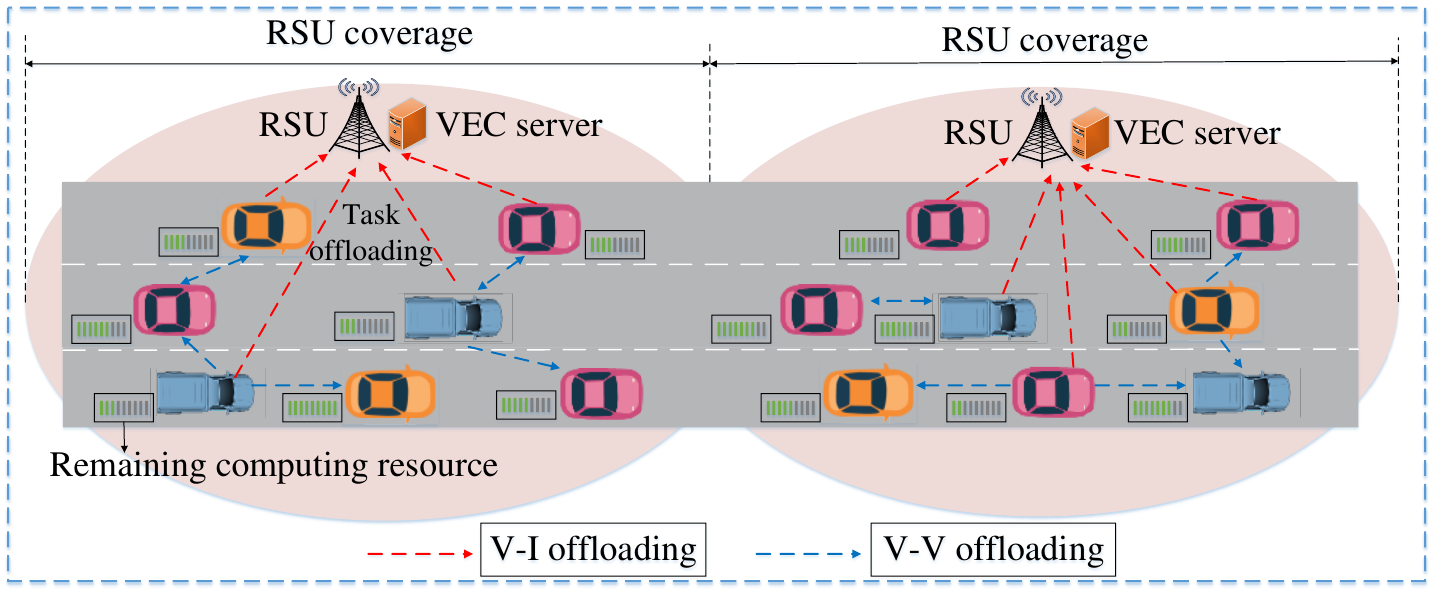}
	\caption{The architecture of computing task offloading in VEC network}
	\label{fig:architecture}
\end{figure}

In this paper, we consider the scenario of the computing task ofﬂoading of vehicles on urban roads in a VEC network, as shown in Fig. \ref{fig:architecture}. RSU is located along the roadside at some distance, and the coverage areas of the adjacent RSUs do not overlap. Thus, according to the coverage areas of RSUs, the whole road can be divided into several adjacent segments, where the vehicle can only communicate with the RSU within its current road segment. When the vehicle crosses the boundary of two road segments, a handover occurs between two adjacent RSUs to finish the message delivery. Each RSU is equipped with a VEC server whose powerful computation capacity can help vehicles quickly handle computing tasks to guarantee their task delay requirements. In this paper, the computing task of the vehicle terminal can be offloaded to the VEC server by vehicle-to- infrastructure (V2I) communication, the neighboring vehicles by V2V communication and the local terminal, respectively.

\begin{figure*}[!t]
	\centering
	\includegraphics[width=14cm]{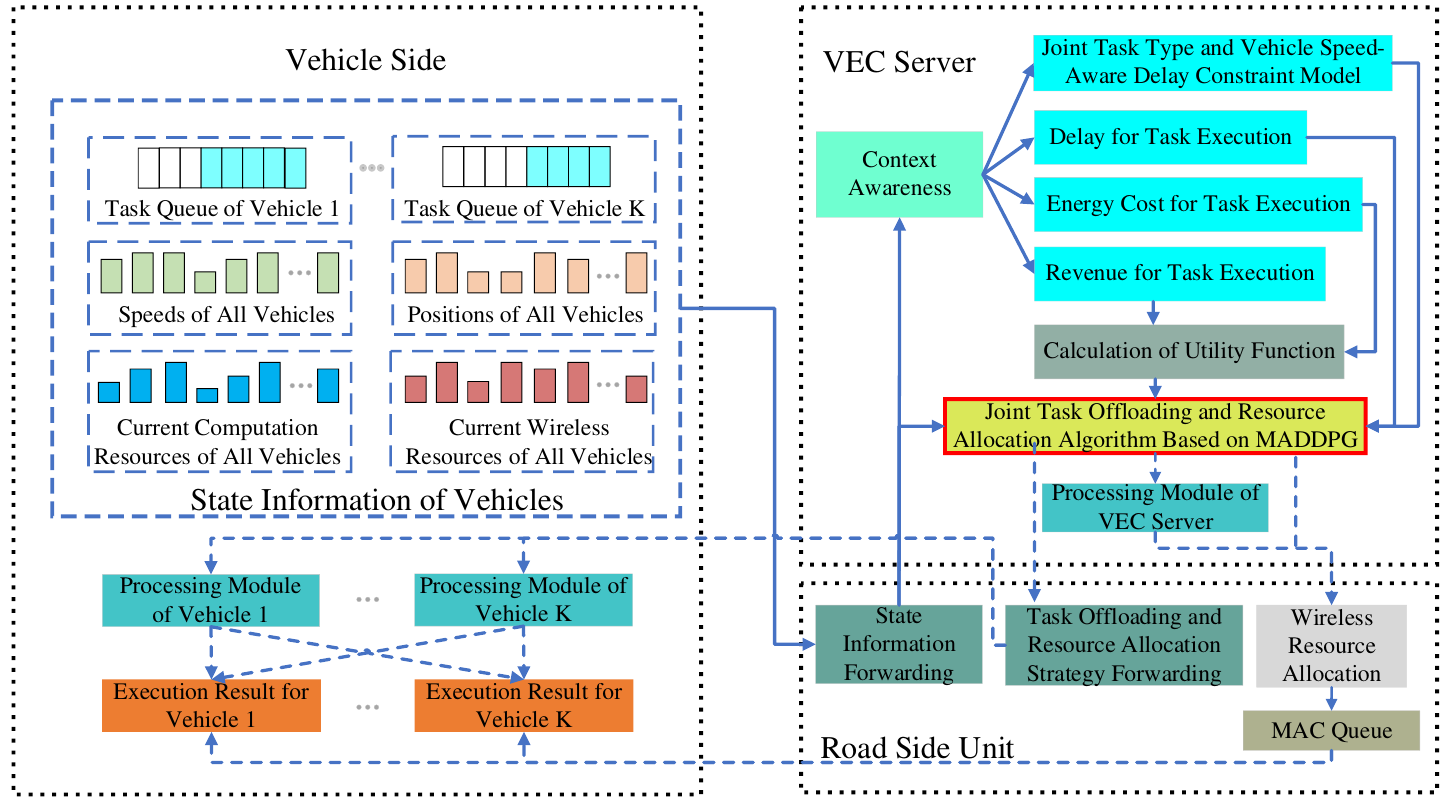}
	\caption{System Framework}
	\label{fig:framework}
\end{figure*}

The VEC server can obtain the state information of the vehicle, i.e., task queue, vehicle speed, position, current computation and wireless resources through the wireless link between the RSU and the vehicle in real time. Due to its powerful computing capability, the VEC server is selected as the central processing module to make decisions on task offloading and resource allocation (including wireless resource allocation and computation resource allocation) strategies, as shown in Fig. \ref{fig:framework}. First, the state information of the vehicle terminals is periodically reported to RSU, which then forwards the received state information to the VEC server. Based on the collected state information of the vehicle terminals, the joint task type and vehicle speed-aware delay constraint model is established to depict the delay constraints of different types of tasks under different vehicle speeds. In addition, the energy cost model and the revenue model of different offloading patterns are established to formulate a utility function model of task offloading and resource allocation decisions. Then, with the objective of maximizing the utility function of all vehicles subject to task delay constraints, the joint task offloading and resource allocation algorithm based on MADDPG is performed in the VEC server to obtain the task offloading and resource allocation strategy. According to the task offloading strategy, when computing tasks are executed locally or on other vehicle terminals, the VEC server will send the corresponding computation and wireless resource allocation strategy to vehicle terminals through the RSU. Otherwise, when computing tasks are executed on the VEC server, the VEC server will send the task processing result to the RSU and then forward to the corresponding vehicle terminal based on the obtained wireless resource allocation strategy.

\subsection{Joint Task Type and Vehicle Speed-aware Delay Constraint Model}
According to the delay tolerance of the in-vehicle computing tasks, computing tasks can be divided into three types: critical application (CA),  high-priority application (HPA) and low-priority application (LPA) \cite{Dziyauddin}. We denote these three types by ${{\phi }_{1}}$, ${{\phi }_{2}}$ and ${{\phi }_{3}}$, respectively. CA task generally refers to the autonomous driving and road safety applications, which need ultra-delay to ensure the safety of driving the vehicle \cite{Gharaibeh}. Therefore, this type of task should be executed locally, and the delay threshold is set to $Th{{r}_{1}}$. The HPA tasks mainly involve image-assisted navigation, parking navigation and some optional security applications \cite{Peng}. The delay tolerance of an HPA task is related to the current vehicle speed, which means that the vehicle with low speed can tolerate a relatively high delay compared with the vehicle with high speed. The delay threshold of HPA task is set to $Th{{r}_{2}}$ when the vehicle speed reaches the maximum road speed limit ${{v}_{\max }}$. LPA tasks generally include multimedia and passenger entertainment activities, so the requirement for delay threshold is relatively slack compared with the previous CA task and HPA task \cite{Feng}. The delay threshold is set to $Th{{r}_{3}}$. In conclusion, different types of tasks demand various delay requirements and computing capability. For an HPA task, it is critical to explore the impact of vehicle speed on task delay. 

\textbf{\textit{Analysis of the impact of task type and vehicle speed on task delay constraint:}} In this paper, we assume that the generated computing tasks are independent task sequences. The computing task of the vehicle $k$ at time $t$ to be processed is defined as ${{\mathcal{I}}_{t,k}}$. Since the delay constraint of an HPA task is related to the current vehicle speed, we mainly analyze the impact of vehicle speed on the task delay constraint of an HPA task. For an HPA task, when the speed of the vehicle is low, the delay threshold of the computing task can be relatively high. With the increase in the speed, the information of the vehicle received from the surrounding environment will increase rapidly because of the longer distance traveled. In this case, the time required to finish the computing task will decrease rapidly. When the speed reaches a higher level, with the increase of speed, the increased amplitude of the information of the vehicle received from the surrounding environment will gradually decrease. In this case, the time required to finish the computing task will decrease slowly.

\begin{figure}[!h]
	\centering
	\includegraphics[width=7cm]{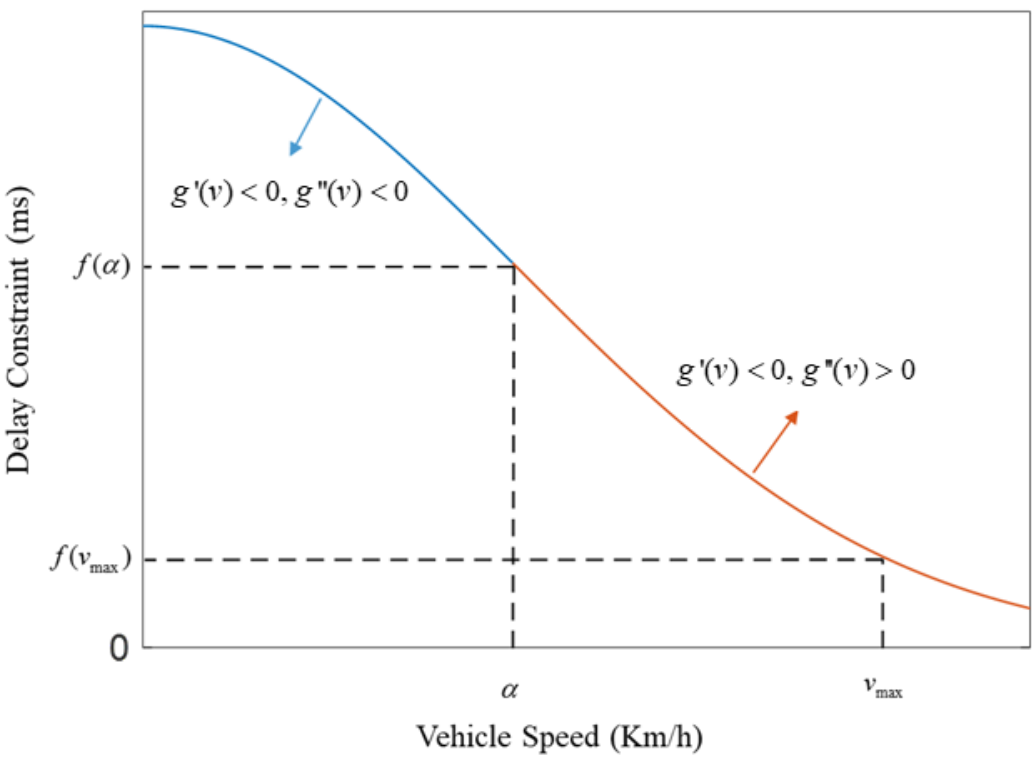}
	\caption{Vehicle speed and delay constraint model for HPA task}
	\label{fig:speed}
\end{figure}

According to the above analysis, we employ a one-tailed normal function, $g(v)$, to reflect the relationship between the vehicle speed and the task delay constraint. The properties of the one-tailed normal function are shown in Fig. \ref{fig:speed}. 

When the vehicle speed is within the range of $(0,\alpha )$, the function satisfies $g'(v)<0, g''(v)<0$, and the decreasing amplitude of the delay constraint gradually grows. When the vehicle speed is within the range of $(\alpha, {{v}_{\max }})$, the function satisfies $g'(v)<0, g''(v)>0$, and the decreasing amplitude of the delay constraint gradually narrows. The properties of this function satisfy the relationship between the vehicle speed and the delay threshold, as previously analyzed. Therefore, it is suitable to use this function to describe the vehicle speed-aware delay constraint model for an HPA task as follows:
\begin{equation}
\begin{aligned}
 \mathcal{T}({{v}_{{{\mathcal{I}}_{t,k}}}})&=Th{{r}_{2}}\frac{1}{\sqrt{2\pi }\alpha }\exp (-\frac{v_{{{\mathcal{I}}_{t,k}}}^{2}}{2{{\alpha }^{2}}})/(\frac{1}{\sqrt{2\pi }\alpha }\exp (-\frac{{{v}_{{{\max }^{2}}}}}{2{{\alpha }^{2}}})) \\ 
& =\exp (-\frac{v_{{{\mathcal{I}}_{t,k}}}^{2}-{{v}_{{{\max }^{2}}}}}{2{{\alpha }^{2}}})Th{{r}_{2}},\text{ }if\text{  }{{\mathcal{I}}_{t,k}}\in {{\phi }_{2}} \\ 
\end{aligned}  
\end{equation}
where ${{v}_{{{\mathcal{I}}_{t,k}}}}$ is the current vehicle speed, and ${{\alpha }^{2}}$ is the variance of the one-tailed normal function. To ensure that the probability that vehicle speed is within the maximum speed exceeds 95\%, we denote $\alpha ={{v}_{\max }}/1.96$. To unify the expression of the delay threshold for different types of tasks, we employ $\Upsilon ({{v}_{{{\mathcal{I}}_{t,k}}}})$ to represent the joint task type and vehicle speed-aware delay constraint model for the computing task ${{\mathcal{I}}_{t,k}}$, as follows: 
\begin{equation}
\Upsilon ({{v}_{{{\mathcal{I}}_{t,k}}}})=\left\{ \begin{aligned}
& Th{{r}_{1}},\quad\quad\;\; if \quad {{\mathcal{I}}_{t,k}}\in {{\phi }_{1}} \\ 
& \mathcal{T}({{v}_{{{\mathcal{I}}_{t,k}}}}),\quad if \quad {{\mathcal{I}}_{t,k}}\in {{\phi }_{2}} \\ 
& Th{{r}_{3}},\quad\quad if \quad {{\mathcal{I}}_{t,k}}\in {{\phi }_{3}} \\ 
\end{aligned} \right.
\end{equation}

\section{Delay, Energy Cost and Revenue Model of Different Offloading Patterns}
For the in-vehicle computing task, raw data are not allowed to be offloaded to neighboring VEC servers through the handover of RSUs because large amounts of raw data transmission between RSUs may result in task offloading failure \cite{Zhang}. In addition, to relieve the traffic and computing burden of computing tasks offloaded to the VEC server, the vehicle terminal with available computation resources can help other vehicles to process computing tasks and obtain the revenue by charging processing fees. Therefore, for generated HPA and LPA tasks at time $t$, there are usually four ways to handle these tasks such as hold on, offloaded to the VEC server, offloaded to other vehicles and local execution, as shown in Fig. \ref{fig:delay}. 

\begin{figure}[!h]
	\centering
	\includegraphics[width=8.8cm]{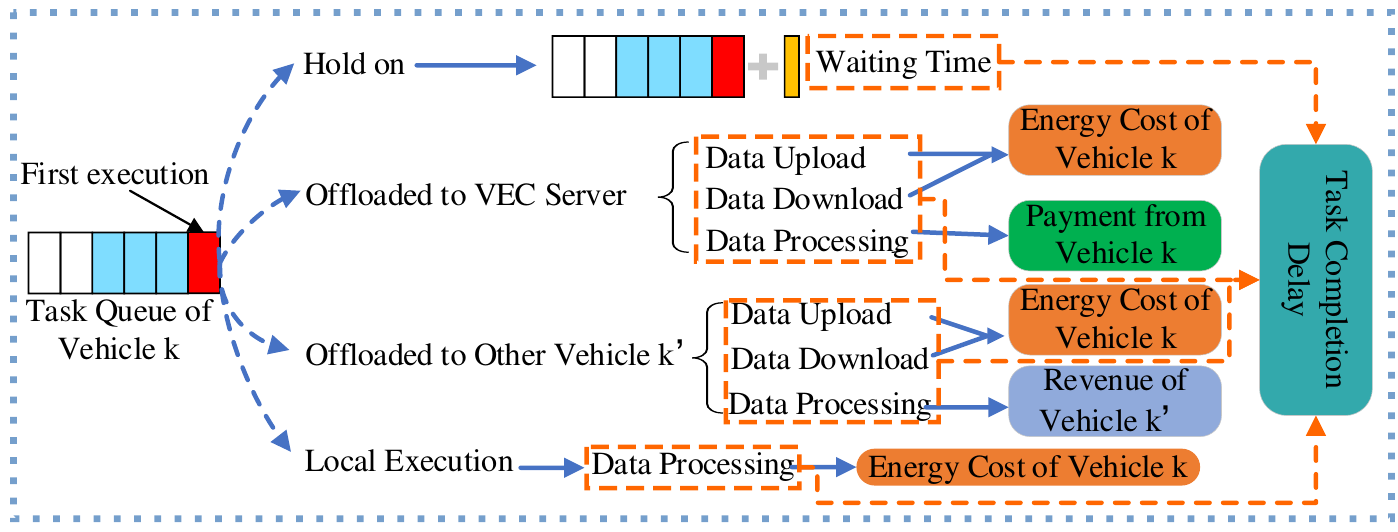}
	\caption{Delay, Energy Cost and Revenue of Different Offloading Patterns}
	\label{fig:delay}
\end{figure}

For the computing task in the task queue, if the current remaining computation resources of the vehicle and the VEC server or the system wireless resources are insufficient to offload and process the computing task, the computing task of the vehicle can choose to wait for a certain time until computation and wireless resources are released. When the computing task of the vehicle is offloaded to the VEC server, task completion delay contains the time spent on data upload, data download and data processing. In this procedure, data upload and data download will consume the energy of the vehicle while data processing needs the vehicle to pay for the VEC server. When the computing task of the vehicle is offloaded to other vehicle $k’$, the local vehicle can establish the wireless link with the surrounding vehicles within 200 m through V2V communication to decrease channel fading effects and guarantee the probability of successful transmission \cite{Ren}. When the computing task of the vehicle is executed locally, task completion delay depends only on the data processing time, and the local vehicle terminal will consume energy. 

\subsection{Delay Model of Task Execution for Different Offloading Patterns}
\subsubsection{Offloading position: VEC server}
For the VEC server, we denote the set of vehicles in its service area by $\mathbb{Q}$, and the number of these vehicles is $K$. When the computing task ${{\mathcal{I}}_{t,k}}$ that belongs to task type $\phi_2$ or $\phi_3$ is offloaded to the VEC server, the task completion time contains the time spent on data upload, data processing and data download. For the task type $\phi_2$, since the size of the output file is much smaller than the size of the input file, the download time can be ignored. Considering that data upload, data processing and data download cannot be executed simultaneously within one transport time interval (TTI), the consumed time, $T{{R}_{{{\mathcal{I}}_{t,k}}}}$, needs to round up, which can be described as:
\begin{equation}
TR_{{{\mathcal{I}}_{t,k}}}^{{}}=\left\{ \begin{aligned}
& \left\lceil \frac{{{c}_{{{\mathcal{I}}_{t,k}}}}}{rv_{k}^{{}}} \right\rceil +\left\lceil \frac{{{\kappa }_{{{\mathcal{I}}_{t,k}}}}{{c}_{{{\mathcal{I}}_{t,k}}}}}{b_{{{\mathcal{I}}_{t,k}}}^{V}{{f}^{V}}} \right\rceil ,\text{  }if\text{ }{{\mathcal{I}}_{t,k}}\in {{\phi }_{2}} \\ 
& \left\lceil \frac{{{c}_{{{\mathcal{I}}_{t,k}}}}}{rv_{k}^{{}}} \right\rceil +\left\lceil \frac{{{\kappa }_{{{\mathcal{I}}_{t,k}}}}{{c}_{{{\mathcal{I}}_{t,k}}}}}{b_{{{\mathcal{I}}_{t,k}}}^{V}{{f}^{V}}} \right\rceil +\left\lceil \frac{{{e}_{{{\mathcal{I}}_{t,k}}}}{{c}_{{{\mathcal{I}}_{t,k}}}}}{\widetilde{rv}_{k}^{{}}} \right\rceil ,\\&\text{ }if\text{ }{{\mathcal{I}}_{t,k}}\in {{\phi }_{3}}
\end{aligned} \right.
\end{equation}
where ${{c}_{{{\mathcal{I}}_{t,k}}}}$ is the file size of task ${{\mathcal{I}}_{t,k}}$ and ${{\kappa }_{{{\mathcal{I}}_{t,k}}}}$ is the calculation density of processing the task ${{\mathcal{I}}_{t,k}}$. ${{e}_{{{\mathcal{I}}_{t,k}}}}$ is the scaling ratio of the downloaded task size relative to the uploaded task size. $b_{{{\mathcal{I}}_{t,k}}}^{V}$ indicates the proportion of the computing resources allocated to the task ${{\mathcal{I}}_{t,k}}$ by the VEC server. ${{f}^{V}}$ denotes the CPU frequency of the VEC server. The transmission capacity between the vehicle and the VEC server can be obtained from the number of allocated channels, channel bandwidth, transmission power and noise power \cite{Huang}. Denote ${{B}^{V}}$ the total available uplink bandwidth for the communication between the VEC server and vehicles. ${{N}^{V}}$ denotes the number of corresponding total uplink channels and $\mathbb{N}_{{}}^{V}$ indicates the corresponding uplink channel set. $\sigma^2$ denotes noise power and $P_{k}^{V}$ is transmission power of vehicle k when the task is offloaded to the VEC server. $h_{k,n}^{V}$ and$I_{k,n}^{V}$ denote the channel gain and interference of uplink channel $n$ between vehicle $k$ and the VEC server, respectively. Then, the available transmission capacity of uplink channel $n$ allocated to vehicle $k$, $rv_{k,n}^{{}}$, can be expressed as    
\begin{equation}
rv_{k,n}^{{}}={{\omega }^{V}}{{\log }_{2}}\left( 1+\frac{P_{k}^{V}\cdot h_{k,n}^{V}}{{{\sigma }^{2}}+I_{k,n}^{V}} \right),\text{  }for\text{  }n\in {{\mathbb{N}}^{V}}
\end{equation}
where $\omega _{{}}^{V}=B_{{}}^{V}/N_{{}}^{V}$. To obtain the uplink transmission capacity of vehicle $k$, we employ $z_{k,n}^{V}$ to indicate whether the uplink channel $n$ is allocated to the vehicle $k$. If it is allocated, $z_{k,n}^{V}=1$, otherwise, $z_{k,n}^{V}=0$. Then, the uplink transmission capacity between vehicle $k$ and VEC server, $rv_{k}^{{}}$, can be depicted as
\begin{equation}
rv_{k}^{{}}=\sum\limits_{n\in \mathbb{N}_{{}}^{V}}{z_{k,n}^{V}rv_{k,n}^{{}}}  
\end{equation}
                       
Similarly, the available transmission capacity of downlink channel $n$ allocated to vehicle $k$, $\widetilde{rv}_{k,n}^{{}}$, can be expressed as
\begin{equation}
\widetilde{rv}_{k,n}^{{}}=\widetilde{\omega }_{{}}^{V}{{\log }_{2}}\left( 1+\frac{P_{k}^{V}\cdot \widetilde{h}_{k,n}^{V}}{{{\sigma }^{2}}+\widetilde{I}_{k,n}^{V}} \right),\text{  }for\text{  }n\in \widetilde{\mathbb{N}}_{{}}^{V}
\end{equation}
where $\widetilde{\omega }_{{}}^{V}=\widetilde{B}_{{}}^{V}/\widetilde{N}_{{}}^{V}$. $\tilde{B}_{{}}^{V}$ is the total available downlink bandwidth for the communication between the VEC server and vehicles. $\tilde{N}_{{}}^{V}$ denotes the number of corresponding total downlink channels and $\widetilde{\mathbb{N}}_{{}}^{V}$ indicates the corresponding downlink channel set. $\tilde{h}_{k,n}^{V}$ and $\tilde{I}_{k,n}^{V}$ denote the channel gain and interference of downlink channel $n$ between vehicle $k$ and VEC server, respectively. Then, the downlink transmission capacity between vehicle $k$ and VEC server, $\widetilde{rv}_{k}^{{}}$, can be depicted as
\begin{equation}
\widetilde{rv}_{k}^{{}}=\sum\limits_{n\in {{\widetilde{\mathbb{N}}}^{V}}}{\tilde{z}_{k,n}^{V}\widetilde{rv}_{k,n}^{{}}}
\end{equation}
where $\tilde{z}_{k,n}^{V}$ indicates whether the downlink channel $n$ is allocated to the vehicle $k$. If it is allocated, $\tilde{z}_{k,n}^{V}=1$, otherwise, $\tilde{z}_{k,n}^{V}=0$. Therefore, the completion delay of task ${{\mathcal{I}}_{t}}(k)$ that is offloaded to the VEC server can be determined based on the allocated computation resources and wireless resources. 

\subsubsection{Offloading position: other vehicles}
To reduce the traffic and computing burden on the VEC server, the computing capability of the vehicle itself is fully utilized to process the computing task generated by other vehicles by V2V communication when computation resources and wireless resources are available. In this scenario, mmwave is employed to realize V2V communication because of its ultra-low delay and fast transmission speed. Denote the set of neighboring vehicles to vehicle k by ${{\mathbb{C}}_{k}}$, whose number is ${{M}_{k}}$. When the computing task ${{\mathcal{I}}_{t,k}}$ belonging to task type $\phi_2$ or $\phi_3$ offloaded to another vehicle $k'$, the task completion time contains data upload, data processing and data download. The task completion time $TV_{{{\mathcal{I}}_{t,k}}}^{k'}$ can be described as:
\begin{equation}\small
TV_{{{\mathcal{I}}_{t,k}}}^{k'}=\left\{ \begin{aligned}
& \left\lceil \frac{{{c}_{{{\mathcal{I}}_{t,k}}}}}{rc_{k}^{k'}} \right\rceil +\left\lceil \frac{{{\kappa }_{{{\mathcal{I}}_{t,k}}}}{{c}_{{{\mathcal{I}}_{t,k}}}}}{b_{{{\mathcal{I}}_{t,k}}}^{k'}{{f}^{k'}}} \right\rceil ,\text{  }if\text{ }{{\mathcal{I}}_{t,k}}\in {{\phi }_{2}},for\text{  }k'\in {{\mathbb{C}}_{k}} \\ 
& \left\lceil \frac{{{c}_{{{\mathcal{I}}_{t,k}}}}}{rc_{k}^{k'}} \right\rceil +\left\lceil \frac{{{\kappa }_{{{\mathcal{I}}_{t,k}}}}{{c}_{{{\mathcal{I}}_{t,k}}}}}{b_{{{\mathcal{I}}_{t,k}}}^{k'}{{f}^{k'}}} \right\rceil +\left\lceil \frac{{{e}_{{{\mathcal{I}}_{t,k}}}}{{c}_{{{\mathcal{I}}_{t,k}}}}}{\widetilde{rc}_{k}^{k'}} \right\rceil ,\\&\text{ }if\text{ }{{\mathcal{I}}_{t,k}}\in {{\phi }_{3}},for\text{  }k'\in {{\mathbb{C}}_{k}} \\ 
\end{aligned} \right.
\end{equation}
where $b_{{{\mathcal{I}}_{t,k}}}^{k'}$ indicates the proportion of computing resources allocated to the task ${{\mathcal{I}}_{t,k}}$ by vehicle  . Let $B_{{}}^{v2v}$ denote the total available uplink bandwidth for the communication between vehicles. ${{N}^{v2v}}$ denotes the number of corresponding total uplink channels, and $\mathbb{N}_{{}}^{v2v}$ indicates the corresponding uplink channel set. $P_{k}^{k'}$ is the transmission power of vehicle $k$ when the task is offloaded to vehicle $k'$. $h_{k,n}^{k'}$ and $I_{k,n}^{k'}$ denote the channel gain and interference of downlink channel $n$ between vehicle $k$ and vehicle$k'$, respectively. Then, the available transmission capacity of uplink channel $n$ allocated to vehicle $k$ for the communication between vehicle $k$ and vehicle $k'$, $rc_{k,n}^{k'}$, can be expressed as:
\begin{equation}
\begin{aligned}
rc_{k,n}^{k'}=&(1-\ell _{k}^{k'})\omega _{{}}^{v2v}{{\log }_{2}}\left( 1+\frac{P_{k}^{k'}\cdot h_{k,n}^{k'}}{{{\sigma }^{2}}+I_{k,n}^{k'}} \right),\\&\text{  }for\text{  }n\in \mathbb{N}_{{}}^{v2v},k'\in {{\mathbb{C}}_{k}}
\end{aligned}
\end{equation}
where $\omega _{{}}^{v2v}=B_{{}}^{v2v}/N_{{}}^{v2v}$. $\ell _{k}^{k'}$ is the alignment delay of beamforming between vehicle $k$ and $k'$, which is related to the relative position of the antenna between vehicles \cite{Elbamby}. Let $z_{k,n}^{k'}$ indicate whether the uplink channel $n$ is allocated to vehicle $k$for the communication between vehicle $k$ and vehicle $k'$. If it is allocated, $z_{k,n}^{k'}=1$, otherwise, $z_{k,n}^{k'}=0$. Then, the uplink transmission capacity between vehicle $k$ and $k'$ can be depicted as
\begin{equation}
rc_{k}^{k'}=\sum\limits_{n\in \mathbb{N}_{{}}^{v2v}}{z_{k,n}^{k'}rc_{k,n}^{k'}},for\text{  }k'\in {{\mathbb{C}}_{k}}
\end{equation}

Similarly, the available transmission capacity of downlink channel $n$ allocated to vehicle $k$ for the communication between vehicle $k$ and vehicle$k'$, $\widetilde{rc}_{k,n}^{k'}$, can be expressed as
\begin{equation}
\begin{aligned}
\widetilde{rc}_{k,n}^{k'}=&(1-\ell _{k}^{k'})\widetilde{\omega }_{{}}^{v2v}{{\log }_{2}}\left( 1+\frac{P_{k}^{k'}\cdot \widetilde{h}_{k,n}^{k'}}{{{\sigma }^{2}}+\widetilde{I}_{k,n}^{k'}} \right),\\&\text{  }for\text{  }n\in \widetilde{\mathbb{N}}_{{}}^{v2v},k'\in {{\mathbb{C}}_{k}}
\end{aligned}
\end{equation}
where $\widetilde{\omega }_{{}}^{v2v}=\widetilde{B}_{{}}^{v2v}/\widetilde{N}_{{}}^{v2v}$. $\tilde{B}_{{}}^{v2v}$ is the total available downlink bandwidth for the communication between vehicles. $\tilde{N}_{{}}^{v2v}$ denotes the number of corresponding total downlink channels and $\widetilde{\mathbb{N}}_{{}}^{v2v}$ indicates the corresponding downlink channel set. $\tilde{h}_{k,n}^{k'}$ and $\tilde{I}_{k,n}^{k'}$ denote the channel gain and interference of downlink channel $n$ between vehicle $k$ and vehicle $k'$, respectively. Let $\tilde{z}_{k,n}^{k'}$ indicate whether the downlink channel $n$ is allocated to vehicle $k$ for the communication between vehicle $k$ and vehicle $k'$. If it is allocated, $\tilde{z}_{k,n}^{k'}=1$, otherwise, $\tilde{z}_{k,n}^{k'}=0$. Then the downlink transmission capacity between vehicle $k$ and $k'$ can be depicted as
\begin{equation}
\widetilde{rc}_{k}^{k'}=\sum\limits_{n\in \widetilde{\mathbb{N}}_{{}}^{v2v}}{\tilde{z}_{k,n}^{k'}\widetilde{rc}_{k,n}^{k'}},for\text{  }k'\in {{\mathbb{C}}_{k}}
\end{equation}
Therefore, the completion delay of task ${{\mathcal{I}}_{t}}(k)$ that is offloaded to other vehicle $k'$ can be determined based on the allocated computation resources and wireless resources.

\subsubsection{Offloading position: local execution}
Since the local terminal of the vehicle owns the processing capacity, a computing task can also be executed locally to save the data uploading and downloading time. For computing task ${{\mathcal{I}}_{t,k}}$ executed locally, the consumed time $T{{L}_{{{\mathcal{I}}_{t,k}}}}$ can be expressed as:
\begin{equation}
TL_{{{\mathcal{I}}_{t,k}}}^{{}}=\left\lceil \frac{{{\kappa }_{{{\mathcal{I}}_{t,k}}}}{{c}_{{{\mathcal{I}}_{t,k}}}}}{b_{{{\mathcal{I}}_{t,k}}}^{k}{{f}^{k}}} \right\rceil
\end{equation}           
where $b_{{{\mathcal{I}}_{t,k}}}^{k}$ is the proportion of computing resources allocated to the task ${{\mathcal{I}}_{t,k}}$ by local vehicle terminal. ${{f}^{k}}$ is the CPU frequency of vehicle $k$. 
Based on above analysis, at time $t$, computing task ${{\mathcal{I}}_{t,k}}$ can select to hold on, be offloaded to the VEC server, be offloaded to other vehicles or be executed locally. Therefore, for computing task ${{\mathcal{I}}_{t,k}}$, the total delay $D({{\mathcal{I}}_{t,k}})$ from generation to completion is derived by
\begin{equation}
\begin{aligned}
D({{\mathcal{I}}_{t,k}})&=t-t_{{{\mathcal{I}}_{t,k}}}^{g}+\tau _{{{\mathcal{I}}_{t,k}}}^{H}{{T}_{h}}+(1-\tau _{{{\mathcal{I}}_{t,k}}}^{H})(\tau _{{{\mathcal{I}}_{t,k}}}^{V}TR_{{{\mathcal{I}}_{t,k}}}^{{}} \\ 
&+ \sum\limits_{k'\in {{\mathbb{C}}_{k}}}{\tau _{{{\mathcal{I}}_{t,k}}}^{k'}TV_{{{\mathcal{I}}_{t,k}}}^{k'}}+\tau _{{{\mathcal{I}}_{t,k}}}^{k}T{{L}_{{{\mathcal{I}}_{t,k}}}}) \\ 
\end{aligned}
\end{equation}
where $t_{k,i}^{g}$ is the generated time of task ${{\mathcal{I}}_{t,k}}$. $\tau _{{{\mathcal{I}}_{t,k}}}^{H}$ indicates whether task ${{\mathcal{I}}_{t,k}}$ holds on. If it holds on, ${{\mathcal{I}}_{t,k}}=1$; otherwise, ${{\mathcal{I}}_{t,k}}=0$. ${{T}_{h}}$ denotes the waiting time. $\tau _{{{\mathcal{I}}_{t,k}}}^{V}$ indicates whether task ${{\mathcal{I}}_{t,k}}$ is offloaded to the VEC server. If it is offloaded, $\tau _{{{\mathcal{I}}_{t,k}}}^{V}=1$; otherwise, $\tau _{{{\mathcal{I}}_{t,k}}}^{V}=0$. $\tau _{{{\mathcal{I}}_{t,k}}}^{k'}$ indicates whether task ${{\mathcal{I}}_{t,k}}$ is offloaded to vehicle $k'$. If it is offloaded, $\tau _{{{\mathcal{I}}_{t,k}}}^{k'}=1$; otherwise, $\tau _{{{\mathcal{I}}_{t,k}}}^{k'}=0$. $\tau _{{{\mathcal{I}}_{t,k}}}^{k'}$ indicates whether task ${{\mathcal{I}}_{t,k}}$ is executed locally. If it is executed locally, $\tau _{{{\mathcal{I}}_{t,k}}}^{k}=1$; otherwise, $\tau _{{{\mathcal{I}}_{t,k}}}^{k}=0$.

\subsection{Energy Cost Model of Task Execution for Different Offloading Patterns}
\subsubsection{Offloading position: VEC server}
When the computing task is offloaded to the VEC server, the energy consumption of the vehicle terminal originates from uploading and downloading the computing task. Therefore, the energy consumption $E{{R}_{{{\mathcal{I}}_{t,k}}}}$ of task ${{\mathcal{I}}_{t,k}}$ offloaded to the VEC server can be depicted as:
\begin{equation}
ER_{{{\mathcal{I}}_{t,k}}}^{{}}=\left\{ \begin{aligned}
& P\frac{{{c}_{{{\mathcal{I}}_{t,k}}}}}{rv_{k}^{{}}},\text{  }if\text{  }{{\mathcal{I}}_{t,k}}\in {{\phi }_{2}} \\ 
& P(\frac{{{c}_{{{\mathcal{I}}_{t,k}}}}}{rv_{k}^{{}}}\text{+}\frac{{{\mu }_{{{\mathcal{I}}_{t,k}}}}{{c}_{{{\mathcal{I}}_{t,k}}}}}{\widetilde{rv}_{k}^{{}}}),\text{  }if\text{  }{{\mathcal{I}}_{t,k}}\in {{\phi }_{3}} \\ 
\end{aligned} \right.
\end{equation}

\subsubsection{Offloading position: other vehicles}
Similarly, when the computing task is offloaded to neighboring vehicle $k'$, the energy consumption of the vehicle terminal originates from uploading and downloading the computing task. Therefore, the energy consumption $EV_{{{\mathcal{I}}_{t,k}}}^{k'}$ of task ${{\mathcal{I}}_{t,k}}$ offloaded to vehicle $k'$ can be depicted as:
\begin{equation}\small
EV_{{{\mathcal{I}}_{t,k}}}^{k'}=\left\{ \begin{aligned}
& P\frac{{{\kappa }_{{{\mathcal{I}}_{t,k}}}}{{c}_{{{\mathcal{I}}_{t,k}}}}}{rc_{k}^{k'}},\text{  }if\text{  }{{\mathcal{I}}_{t,k}}\in {{\phi }_{2}},for\text{  }k'\in {{\mathbb{C}}_{k}} \\ 
& P(\frac{{{\kappa }_{{{\mathcal{I}}_{t,k}}}}{{c}_{{{\mathcal{I}}_{t,k}}}}}{rc_{k}^{k'}}\text{+}\frac{{{\mu }_{{{\mathcal{I}}_{t,k}}}}{{c}_{{{\mathcal{I}}_{t,k}}}}}{\widetilde{rc}_{k}^{k'}}),\text{  }if\text{  }{{\mathcal{I}}_{t,k}}\in {{\phi }_{3}},for\text{  }k'\in {{\mathbb{C}}_{k}} \\ 
\end{aligned} \right.
\end{equation}

\subsubsection{Offloading position: local execution}
In addition to offloading tasks to the VEC server and other vehicles, computing task ${{\mathcal{I}}_{t,k}}$ can also be executed locally to save the energy cost of data upload and download. The local consumed energy $E{{L}_{{{\mathcal{I}}_{t,k}}}}$ can be calculated according to the assigned computation resource, as follows:
\begin{equation}
EL_{{{\mathcal{I}}_{t,k}}}^{{}}={{\xi }_{{{\mathcal{I}}_{t,k}}}}{{\kappa }_{{{\mathcal{I}}_{t,k}}}}{{c}_{{{\mathcal{I}}_{t,k}}}}{{(b_{{{\mathcal{I}}_{t,k}}}^{k}{{f}^{k}})}^{2}}
\end{equation}
where ${{\xi }_{{{\mathcal{I}}_{t,k}}}}$ is the energy density of the processing task ${{\mathcal{I}}_{t,k}}$ \cite{Zhan2}.
According to the different offloading positions of computing task ${{\mathcal{I}}_{t}}(k)$, including the VEC server, other vehicle terminals and the local terminal, the consumed energy $E(t)$ of all vehicles served by the VEC server can be derived by
\begin{equation}
\begin{aligned}
{{E}_{t}}&=\sum\limits_{k\in \mathbb{Q}}{(1-\tau _{{{\mathcal{I}}_{t,k}}}^{H})(\tau _{{{\mathcal{I}}_{t,k}}}^{V}E{{R}_{{{\mathcal{I}}_{t,k}}}}+\sum\limits_{k'\in {{\mathbb{C}}_{k}}}{\tau _{{{\mathcal{I}}_{t,k}}}^{k'}EV_{{{\mathcal{I}}_{t,k}}}^{k'}}}\\&+\tau _{{{\mathcal{I}}_{t,k}}}^{k}E{{L}_{{{\mathcal{I}}_{t,k}}}})
\end{aligned}
\end{equation}

\subsection{Revenue Model of Task Execution for Different Offloading Patterns}
For the generated computing tasks of vehicles, the cost of offloading tasks to the VEC server and other vehicles is different. Generally, deploying VEC servers need to pay for land occupation and purchase high-performance equipment, which can bring a high cost for task offloading and processing. However, due to the advantages of V2V communication and the computing capacity of the vehicle itself, vehicles can complete task offloading and process tasks with a low cost. Therefore, the price of offloading to other vehicles to process is lower compared with the price of offloading to the VEC sever. We denote the processing unit price of the VEC server by${{p}_{V}}$, and the processing unit price of other vehicles by${{p}_{v2v}}$, respectively.

For the vehicle $k$ at time $t$, the actual revenue, $T{{R}_{t,k}}$, is related with processing revenue from other vehicles, cost of processing local tasks and payment for offloading the task to the VEC server and other vehicles, which can be expressed as: 
\begin{equation}
\begin{aligned}
T{{R}_{t,k}}&=\sum\limits_{k'\in {{\mathbb{C}}_{k}}}{\tau _{k',{{\mathcal{I}}_{t,k}}}^{k}{{p}_{v2v}}{{c}_{{{\mathcal{I}}_{t,k}}}}}-\tau _{k,{{\mathcal{I}}_{t,k}}}^{V}{{p}_{V}}{{c}_{{{\mathcal{I}}_{t,k}}}} \\&-\sum\limits_{k'\in {{\mathbb{C}}_{k}}}{\tau _{k,{\mathcal{I}}_{t,k}}^{k'}{{p}_{v2v}}{{c}_{{{\mathcal{I}}_{t,k}}}}-}\tau _{k,{{\mathcal{I}}_{t,k}}}^{k}{{p}_{0}}{{c}_{{{\mathcal{I}}_{t,k}}}}
\end{aligned}
\end{equation}
where ${{p}_{0}}$ is the unit price of processing the local task. Therefore, we can obtain the total actual revenue of all vehicles as follows:
\begin{equation}
T{{R}_{t}}=\sum\limits_{k\in \mathbb{Q}}{T{{R}_{t,k}}}
\end{equation}

\section{Delay Constrained Joint Computing Task Offloading and Resource Allocation Optimization Based on Revenue and Energy Cost}

\subsection{Problem Formulation}
To improve the performance of the VEC network, we need to decrease the vehicles' energy cost and enhance the revenue of processing tasks. Therefore, these two factors are utilized to formulate the utility function, ${{U}_{t}}$, which can be expressed as:
\begin{equation}
{{U}_{t}}={{\beta }_{1}}T{{R}_{t}}-{{\beta }_{2}}{{E}_{t}}
\end{equation}
where ${{\beta }_{1}},{{\beta }_{2}}$ are positive values. The problem with the objective of maximizing the overall utility level of the vehicles, subject to the constraints of the task delay threshold, computation resources and wireless resources of the VEC server and vehicle terminals, is as follows:
\begin{equation}
\begin{aligned}
& \underset{{{X}_{t}}(k,k',n),\forall k,k',n}{\mathop{\max }}\,{{U}_{t}} \\ 
& s.t. \\ 
& (c1)\tau _{{{\mathcal{I}}_{t,k}}}^{V}+\sum\limits_{k'\in {{\mathbb{C}}_{k}}}{\tau _{{{\mathcal{I}}_{t,k}}}^{k'}}+\tau _{{{\mathcal{I}}_{t,k}}}^{k}+\tau _{{{\mathcal{I}}_{t,k}}}^{H}\le 1,\forall k\in \mathbb{Q} \\ 
& (c2)b_{{{\mathcal{I}}_{t,k}}}^{k}+\sum\limits_{k'\in {{\mathbb{C}}_{k}}}{b_{{{\mathcal{I}}_{t,k'}}}^{k}}\le 1,\forall k\in \mathbb{Q} \\ 
& (c3)\sum\limits_{k}{b_{{{\mathcal{I}}_{t,k}}}^{V}}\le 1,\forall k\in \mathbb{Q} \\ 
& (c4)\sum\limits_{k}{z_{k,n}^{V}}\le 1,\forall k\in \mathbb{Q},n\in \mathbb{N}_{{}}^{V} \\ 
& (c5)\sum\limits_{k}{\tilde{z}_{k,n}^{V}}\le 1,\forall k\in \mathbb{Q},n\in \widetilde{\mathbb{N}}_{{}}^{V} \\ 
& (c6)\sum\limits_{k}{z_{k,n}^{k'}}\le 1,\forall k\in \mathbb{Q},k'\in {{\mathbb{C}}_{k}},n\in \mathbb{N}_{{}}^{v2v} \\ 
& (c7)\sum\limits_{k}{\tilde{z}_{k,n}^{k'}}\le 1,\forall k\in \mathbb{Q},k'\in {{\mathbb{C}}_{k}},n\in \widetilde{\mathbb{N}}_{{}}^{v2v} \\ 
& (c8)D({{\mathcal{I}}_{t,k}})\le \Upsilon ({{v}_{{{\mathcal{I}}_{t,k}}}}),\forall k\in \mathbb{Q} \\ 
\end{aligned}
\end{equation}
where ${{X}_{t}}(k,k',n)=(\tau _{{{\mathcal{I}}_{t,k}}}^{V},\tau _{{{\mathcal{I}}_{t,k}}}^{k'},\tau _{{{\mathcal{I}}_{t,k}}}^{H},b_{{{\mathcal{I}}_{t,k}}}^{k},b_{{{\mathcal{I}}_{t,k}}}^{k'},z_{k,n}^{V},\tilde{z}_{k,n}^{V},\\z_{k,n}^{k'},\tilde{z}_{k,n}^{k'})$. Constraint (c1) implies that computing task ${{\mathcal{I}}_{t(k)}}$ can be handled by only one way such as being offloaded to the VEC server, other vehicle terminals, executed locally and holding on. Constraint (c2) indicates that the allocated computation resources of vehicle k cannot exceed its own computing capability. Constraint (c3) indicates that the allocated computation capacity of the VEC server cannot exceed its own computing capacity. Constraints (c4) and (c5) indicate that each uplink and downlink channel for the communication between the VEC server and vehicles must be assigned to one and only one vehicle at each scheduling period. Constraints (c6) and (c7) indicate that each uplink and downlink channel for the communication between vehicles must be assigned to one and only one vehicle at each scheduling period. Constraint (c8) indicates that computing task ${{\mathcal{I}}_{t,k}}$ should be completed within the delay threshold.

\subsection{Deep Reinforcement Learning-Based Solution Method}
Equation (22) is a multi-vehicle cooperation and competition problem, which is obviously an NP-hard problem. Therefore, we employ the deep reinforcement learning method to solve the proposed computing task offloading and resource allocation problem. First, we formulate our problem as an MDP to accurately describe the offloading and resource allocation decision process and then utilize the multi-agent deep deterministic policy gradient (MADDPG) \cite{Lowe}\cite{Zhang5} to train. Based on the trained model, we can quickly obtain the near optimal task of offloading and resource allocation policy for each vehicle. In what follows, we will present the elements of MDP, including state space, action space and reward function.

\subsubsection{State Space}
We defined the state space of vehicle $k$ at time $t$ as ${{s}_{t,k}}$, including the state information of vehicle $k$, other vehicles and the VEC server, which is depicted as
\begin{equation}
\begin{aligned}
& {{s}_{t,k}}=[{{v}_{t,1}},...,{{v}_{t,K}},{{d}_{t,1}},...,{{d}_{t,K}},{{c}_{t,1}},...,{{c}_{t,K}},r{{b}_{t,V}},r{{b}_{t,k}}, \\ 
& \hat{\tau }_{t,k}^{H},\hat{\tau }_{t,k}^{V},\hat{\tau }_{t,k}^{1},...,\hat{\tau }_{t,k}^{k'},...,\hat{\tau }_{t,k}^{K},sb_{t,k}^{V},sb_{t,k}^{k},sb_{t,1}^{k},...,sb_{t,k'}^{k},...,\\&sb_{t,K}^{k}, 
sz_{t,k,1}^{V},...,sz_{t,k,N_{{}}^{V}}^{V},\widetilde{sz}_{t,k,1}^{V}...,\widetilde{sz}_{t,k,\widetilde{N}_{{}}^{V}}^{V},sz_{t,k,1}^{1},...,\\&sz_{t,k,n}^{k'},...,sz_{t,k,N_{{}}^{v2v}}^{K},\widetilde{sz}_{t,k,1}^{1},...,\widetilde{sz}_{t,k,n}^{k'},\widetilde{sz}_{t,k,\widetilde{N}_{{}}^{v2v}}^{K}] \\ 
\end{aligned}
\end{equation}
where ${{v}_{t,k}},{{d}_{t,k}},{{c}_{t,k}}$ are vehicle speed, position and file size to be processed for vehicle $k$ at time $t$, respectively. $r{{b}_{t,V}}$ and $r{{b}_{t,k}}$ are the current computation capacity of the VEC server and vehicle $k$ at time $t$, respectively. $\hat{\tau }_{t,k}^{H}$ indicates whether the task of vehicle $k$ is holding on at time $t$. If it is holding on, $\hat{\tau }_{t,k}^{H}=1$; otherwise, $\hat{\tau }_{t,k}^{H}=0$. $\hat{\tau }_{t,k}^{V}$ indicates whether the task of vehicle $k$ has been offloaded to the VEC server at time $t$. If it has been offloaded to the VEC server, $\hat{\tau }_{t,k}^{V}=1$; otherwise, $\hat{\tau }_{t,k}^{V}=0$. $\hat{\tau }_{t,k}^{k'}(1\le k'\le K)$ indicates whether the task of vehicle $k$ has been offloaded to vehicle $k'$ at time $t$. If it has been offloaded to vehicle $k'$, $\hat{\tau }_{t,k}^{k'}=1$; otherwise, $\hat{\tau }_{t,k}^{k'}=0$. $sb_{t,k}^{V}$ is the ratio of computation resources allocated by the VEC sever to vehicle $k$ at time $t$. $sb_{t,k}^{k}$ is the ratio of computation resources allocated by vehicle $k$ to itself at time $t$.$sb_{t,k'}^{k}(1\le k'\le K)$ is the ratio of computation resources allocated by vehicle $k$ to the other vehicle $k'$at time $t$. $sz_{t,k,1}^{V},...,sz_{t,k,{{N}^{V}}}^{V}$ indicates whether the uplink channel resource for the communication between the VEC server and vehicle k is available at time $t$. If it is available, the value is 1; otherwise, the value is 0. $\widetilde{sz}_{t,k,1}^{V},...,\widetilde{sz}_{t,k,{{N}^{V}}}^{V}$ indicates whether the downlink channel resource for the communication between the VEC server and vehicle $k$ is available at time $t$. If it is available, the value is 1; otherwise, the value is 0. $sz_{t,k,n}^{k'}(1\le k'\le K)$ indicates whether the uplink channel $n$ for the communication between vehicles $k$ and $k'$ is available at time $t$. If it is available, the value is 1; otherwise, the value is 0. $\widetilde{sz}_{t,k,n}^{k'}(1\le k'\le K)$ indicates whether the downlink channel $n$ for the communication between vehicle $k$ and $k'$ is available at time $t$. If it is available, the value is 1; otherwise, the value is 0. Therefore, the state space of the system can be defined as: ${{S}_{t}}=({{s}_{t,1}},...{{s}_{t,k}}...,{{s}_{t,K}})$.

\subsubsection{Action Space}
For vehicle $k$ at time $t$, the action space ${{a}_{t,k}}$ contains whether to hold on, offload task to the VEC server and other vehicles, computation resource obtained from VEC server and other vehicles, the uplink and downlink channels allocated to the vehicles, which can be expressed as:
\begin{equation}
\begin{aligned}
a_{t,k}^{{}}&=[\tau _{t,k}^{H},\tau _{t,k}^{V},\tau _{t,k}^{1},...,\tau _{t,k}^{K},b_{t,k}^{V},b_{t,k}^{k},b_{t,1}^{k},...,b_{t,K}^{k}, \\ 
& z_{t,k,1}^{V},...,z_{t,k,{{N}^{V}}}^{V},\tilde{z}_{t,k,1}^{V},...,\tilde{z}_{t,k,{{{\tilde{N}}}^{V}}}^{V}, \\ 
& z_{t,k,1}^{1},...,z_{t,k,n}^{k'},...,z_{t,k,{{N}^{v2v}}}^{K},\tilde{z}_{t,k,1}^{1},...,\tilde{z}_{t,k,n}^{k'},...,\tilde{z}_{t,k,{{{\tilde{N}}}^{v2v}}}^{K}] \\ 
\end{aligned}
\end{equation}
Therefore, the action space of the system can be defined as: ${{A}_{t}}=\{{{a}_{t,1}},...{{a}_{t,k}}...,{{a}_{t,K}}\}$.

\subsubsection{Reward Function}
The goal of this paper is to maximize the overall utility level of vehicles within task delay thresholds, which can be realized by allocating the computation resources of the VEC server and vehicles and system wireless resources. Therefore, we set rewards based on constraint conditions and the objective function to accelerate the training speed. After taking action ${{a}_{t,k}}$, if the state of vehicle $k$ does not satisfy the constraints (c1)-(c7), the reward function is defined as:
\begin{equation}
\begin{aligned}
{{r}_{t,k}}&={{\ell }_{1}}+{{\Gamma }_{1}}\cdot (\hat{\tau }_{t,k}^{V}+\sum\limits_{k'\in {{\mathbb{C}}_{k}}}{\hat{\tau }_{t,k}^{k'}}+\hat{\tau }_{t,k}^{k}+\hat{\tau }_{t,k}^{H}-1)\cdot \\& {{\Lambda }_{(\hat{\tau }_{t,k}^{V}+\sum\limits_{k'\in {{\mathbb{C}}_{k}}}{\hat{\tau }_{t,k}^{k'}}+\hat{\tau }_{t,k}^{k}+\hat{\tau }_{t,k}^{H}\le 1)}} \\ 
& +{{\Gamma }_{2}}\cdot (sb_{t,k}^{k}+\sum\limits_{k'\in {{\mathbb{C}}_{k}}}{sb_{t,k'}^{k}}-1)\cdot {{\Lambda }_{(sb_{t,k}^{k}+\sum\limits_{k'\in {{\mathbb{C}}_{k}}}{sb_{t,k'}^{k}}\le 1)}}\\&+{{\Gamma }_{3}}\cdot (\sum\limits_{k}{sb_{t,k}^{V}}-1)\cdot {{\Lambda }_{(\sum\limits_{k}{sb_{t,k}^{V}}\le 1)}} \\ 
& +{{\Gamma }_{4}}\cdot (\sum\limits_{k}{sz_{t,k,n}^{V}}-1)\cdot {{\Lambda }_{(\sum\limits_{k}{sz_{t,k,n}^{V}}\le 1)}}\\&+{{\Gamma }_{5}}\cdot (\sum\limits_{k}{\widetilde{sz}_{t,k,n}^{V}}-1)\cdot {{\Lambda }_{(\sum\limits_{k}{\widetilde{sz}_{t,k,n}^{V}}\le 1)}} \\ 
& +{{\Gamma }_{6}}\cdot (\sum\limits_{k}{sz_{t,k,n}^{k'}}-1)\cdot {{\Lambda }_{(\sum\limits_{k}{sz_{t,k,n}^{k'}}\le 1)}}\\&+{{\Gamma }_{7}}\cdot (\sum\limits_{k}{\widetilde{sz}_{t,k,n}^{k'}}-1)\cdot {{\Lambda }_{(\sum\limits_{k}{\widetilde{sz}_{t,k,n}^{k'}}\le 1)}} \\ 
\end{aligned}
\end{equation}
where ${{\Lambda }_{(\centerdot )}}$ indicates that if the condition $(\centerdot)$ is not satisfied, the value is -1; otherwise, the value is 0. ${{\ell }_{1}},{{\Gamma }_{1}},{{\Gamma }_{2}},{{\Gamma }_{3}},{{\Gamma }_{4}},{{\Gamma }_{5}},{{\Gamma }_{6}},{{\Gamma }_{7}}$ are experimental parameters. After taking action ${{a}_{t,k}}$, if the state of vehicle $k$ satisfies all constraints (c1)-(c7), the reward function is defined as:
\begin{equation}
{{r}_{t,k}}={{\ell }_{2}}+\exp (\Upsilon ({{v}_{{{\mathcal{I}}_{t,k}}}})-D({{\mathcal{I}}_{t,k}}))
\end{equation}
where ${{\ell }_{2}}$ is the experimental parameter. After taking action ${{a}_{t,k}}$, if the state of vehicle $k$ satisfies all constraints (c1)-(c8), the reward function is defined as:
\begin{equation}
{{r}_{t,k}}={{\ell }_{3}}\text{+}{{\Gamma }_{8}}\cdot \exp ({{U}_{t,k}})
\end{equation}
where ${{\ell }_{3}},{{\Gamma }_{8}}$ are experimental parameters.    

\subsection{Joint Offloading and Resource Allocation Algorithm Based on MADDPG}
\begin{figure}[!h]
	\centering
	\includegraphics[width=8.8cm]{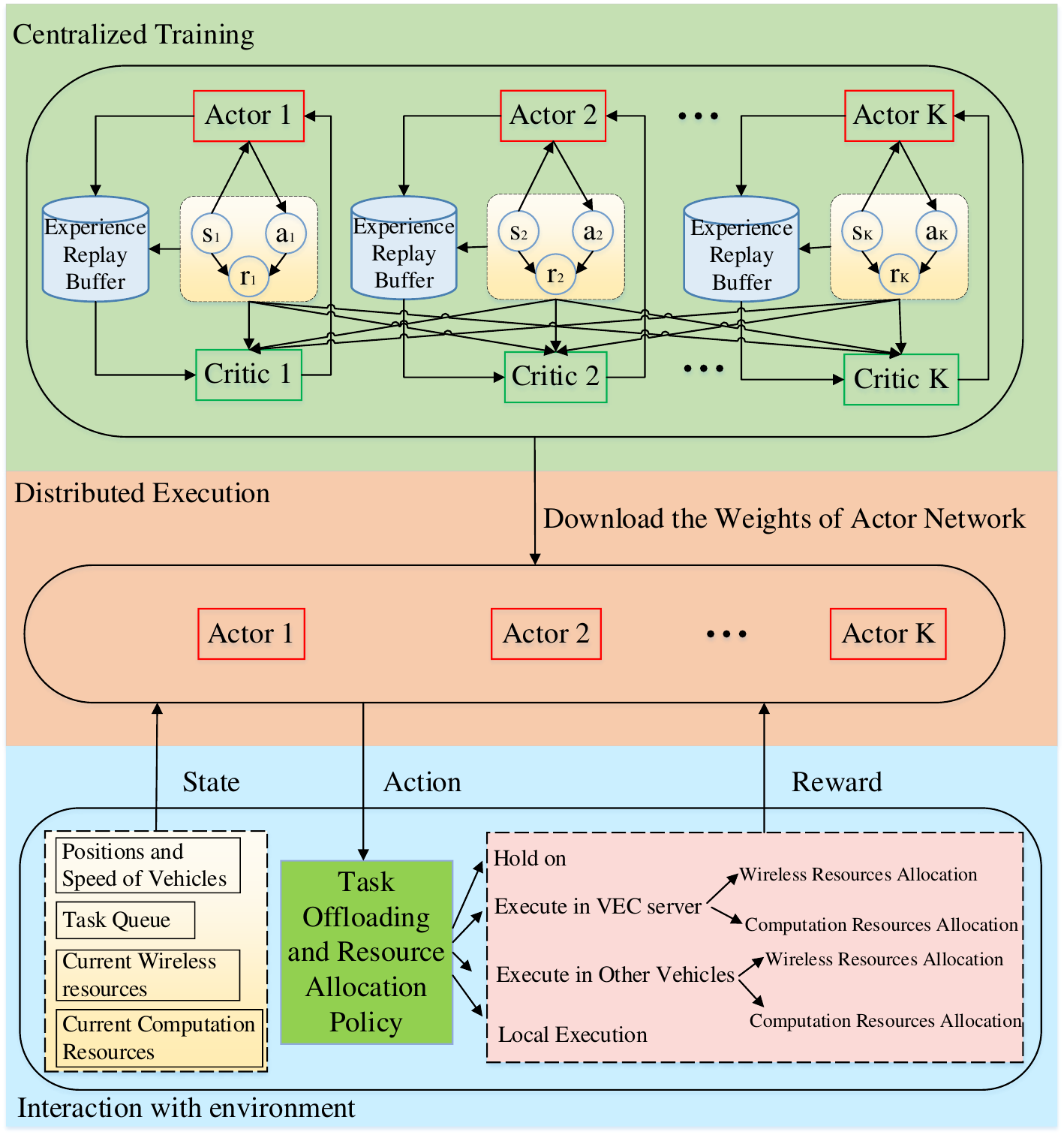}
	\caption{The Proposed JORA-MADDPG Framework}
	\label{fig:MADDPG}
\end{figure}

The proposed JORA-MADDPG framework is shown in Fig. \ref{fig:MADDPG}. Each vehicle is regarded as an agent and owns two networks: an actor network $\pi$ and a critic network Q. The function of the actor network is to calculate the action to be executed based on the state of the agent, while the critic network is responsible for evaluating the action calculated by the actor network. The experience replay buffer is utilized to store a certain number of training data points. When updating the actor network and critic network, the critic network can read the data from the experience replay buffer randomly to break correlations in the training data and make the training process more stable. In the centralized training phase, the actor network obtains observation information only from itself, while the critic network acquires the actions and observations of all agents. In the distributed execution phase, the critic network is not involved, and each agent needs only the actor network. By interacting with the environment, each agent can make task offloading and resource allocation policy.  

The centralized training process is composed of $K$ agents, whose network parameters are $\theta =\{{{\theta }_{1}},...,{{\theta }_{K}}\}$. We denote $\mu =\{{{\mu }_{{{\theta }_{1}}}},...,{{\mu }_{{{\theta }_{K}}}}\}$ (abbreviated as $\mu_i$) as the set of all agent deterministic policies. Therefore, for the deterministic policy ${{\mu }_{k}}$ of agent $k$, the gradient can be depicted as:
\begin{equation}
\begin{aligned}
&{{\nabla }_{{{\theta }_{k}}}}J({{\mu }_{k}})\\&={{\mathbb{E}}_{S,A\sim \mathcal{D}}}[{{\nabla }_{{{\theta }_{k}}}}{{\mu }_{k}}({{a}_{k}}|{{s}_{k}}){{\nabla }_{{{a}_{k}}}}Q_{k}^{\mu }(S,{{a}_{1}},...,{{a}_{K}}){{|}_{{{a}_{k}}={{\mu }_{k}}({{s}_{k}})}}]
\end{aligned}
\end{equation}
where $\mathcal{D}$ is the experience replay buffer, which contains a series of $(S,A,{{S}^{'}},R)$. $Q_{k}^{\mu }(s,{{a}_{1}},...,{{a}_{K}})$ is the Q-value function. For the critic network, it can be updated according to the loss function as follows:
\begin{equation}
\begin{aligned}
& \mathcal{L}({{\theta }_{k}})={{\mathbb{E}}_{S,A,R,{{S}^{'}}}}{{[Q_{k}^{\mu }(S,a_{1}^{{}},...,a_{K}^{{}})-y)}^{2}}] \\ 
& \text{where} \;\; y=r_{k}^{{}}+\gamma Q_{k}^{{{\mu }^{'}}}({{S}^{'}},a_{1}^{'},...,a_{K}^{'}){{|}_{a_{j}^{'}=\mu _{j}^{'}({{s}_{j}})}} \\ 
\end{aligned}
\end{equation}
where $\gamma $ is the discount factor. The action network is updated by minimizing the policy gradient of the agent, which can be expressed as
\begin{equation}
{{\nabla }_{{{\theta }_{k}}}}J\approx \frac{1}{X}\sum\limits_{j}{{{\nabla }_{{{\theta }_{k}}}}}{{\mu }_{k}}(s_{k}^{j}){{\nabla }_{{{a}_{k}}}}Q_{k}^{\mu }({{S}^{j}},a_{1}^{j},...,a_{K}^{j}){{|}_{{{a}_{k}}={{\mu }_{k}}(s_{k}^{j})}}
\end{equation}
where $X$ is the size of mini-batch, $j$ is the index of samples. The specific training algorithm of JORA-MADDPG is shown in Algorithm 1, and the execution algorithm is presented in Algorithm 2.

\begin{algorithm}[!h]
	\caption{Training Algorithm of JORA-MADDPG}
	\label{alg2}
	\textbf{Initialize:} the positions, speed, task queue, computing resources and wireless resources of all vehicles. Initialize the computation and wireless resources of the VEC server. Initialize the weights of actor and critic networks.
	
	\For{$t= 1:M$}
	{
		Initialize a random process $\mathcal{N}$ for action exploration;
		
		Receive initial state $S$;
		
		\For{each vehicle $k=1,...,K$}
		{
			Execute actions ${{a}_{k}}$ and obtain new state $s_{t,k}^{'}$;
			
			\uIf{ the $s_{t,k}^{'}$ does not satisfy constraints (c1)-(c7) in Eq.(22):}
			{
				Obtain the reward of vehicle $k$ based on Eq.(25);
			}
			\uElseIf{the $s_{t,k}^{'}$ satisfy all constraints (c1)-(c7) in Eq.(22):}	 
			{
				Obtain the reward of vehicle $k$ based on Eq.(26);
			}
			\uElseIf{the $s_{t,k}^{'}$ satisfy all constraints (c1)-(c8) in Eq.(22):}
			{
				Obtain the reward of vehicle $k$ based on Eq.(27);
			}
			
			\textbf{end}
			
			Obtain	the action $A$, new state ${{S}^{'}}$ and reward $R$;
			
			Store $(S,A,{{S}^{'}},R)$ in replay buffer $\mathcal{D}$;
			
		}
		\For{each vehicle $k=1,...,K$}
		{
			Sample a random mini-batch of $X$ samples $({{S}^{j}},{{A}^{j}},{{R}^{j}},{{S}^{'}}^{j})$ from $\mathcal{D}$;
			
			Update the critic network by minimizing the loss function, Eq.(29);
			
			Update actor network using the sampled policy gradient, Eq.(30);
		} 
		Update the target network parameters of each vehicle $k$: $\theta _{k}^{'}\leftarrow \delta {{\theta }_{k}}+(1-\delta )\theta _{k}^{'}$
	}
\end{algorithm}

\begin{algorithm}[!h]
	\caption{Execution Algorithm of JORA-MADDPG }
	\label{alg2}
	\textbf{Input:}  the positions, speed, task queue, computation resources and wireless resources of all vehicles. Input the current computation and wireless resources of the VEC server. Input the weights of actor network and execution time $T$.
	
	\textbf{Initialize:} the state space $S$, the action space $A$ and the reward $R$ of all vehicles.
	
	\For{$t= 1:T$}
	{		
		\For{each vehicle $k=1,...,K$}
		{
			Obtain the current state information of vehicle $k$, ${{s}_{t,k}}$;
			
			Execute action ${{a}_{t,k}}$ based on the weights of actor network of vehicle $k$;
			
			Obtain the new state $s_{t,k}^{'}$ and reward ${{r}_{t,k}}$; 
			
			Store action ${{a}_{t,k}}$, state $s_{t,k}^{'}$ and ${{r}_{t,k}}$ into ${{A}_{t}}$, ${{S}_{t}}$ and${{R}_{t}}$, respectively;
			
			\textbf{Return} $A_t$;
		}
	}
\end{algorithm}

\section{Simulation Results}
\subsection{Parameter Setting}
We conduct the experiments on DELL PowerEdge (DELLR940XA, 4*GOLD-5117, RTX2080Ti). We utilize Pytorch 1.5.0 on Ubuntu 18.04.4 LTS in the simulations to implement JORA-MADDPG algorithm and compared it with five compared algorithms. Since the GPU and GPU processing speed are limited, we reduce the computing capacity of the VEC server and the vehicle terminal compared with the actual scenario. And assuming that one VEC server can simultaneously provide computing task processing service for at most 15 vehicles. The time slot is set as 1 ms and the number of training episodes is set as $1.7\times {{10}^{6}}$. The specific simulation parameter configuration is presented in Table \ref{tab:para}. In JORA-MADDPG, our critic network is a five-layer fully connected neural network i.e., one input layer, three hidden layers and one output layer. Tanh function is used in the output layer of actor network to constraint the output action value. The specific parameter settings of neural network and training parameters are given in Table \ref{tab:neu}. The algorithms compared in this section are as follows:

\textbf{All Local Execution (AL):} All computation tasks are executed locally.

\textbf{All VEC Execution (AV):} The CA tasks are executed locally, while HPA and LPA tasks are executed in VEC server. The resource allocation strategy is based on the size of task.

\textbf{Random Offloading (RD):} The HPA and LPA tasks are executed locally, in VEC server and in other vehicles based on the uniform distribution. The resource allocation strategy is based on the size of task.

\textbf{Energy and Delay Greedy (EDG):} The offloading strategy is based on vehicle’s channel state and resource allocation strategy is based on the size of task, in order to decrease the energy cost and execution delay while increasing revenue in each step. 

\textbf{Deep Deterministic Policy Gradient (DDPG)[21]:} The state information of all vehicles and their corresponding local VEC server is formulated into the state space of one agent. The action space consists of offloading strategy and resource allocation strategy for all vehicles. The reward function is designed based on the task completion delay, task processing revenue and energy consumption of vehicles.

\begin{table}[!h]	\centering \scriptsize
	\caption{Simulation Parameter Configuration}
	\begin{tabular}{|c|c|ccc}
		\cline{1-2}
		\bf{Parameter}                                                                  & \bf{Value}                                &  &  &  \\ \cline{1-2}
		Number of vehicles                                                                    & 7 ,9, 11, 13, 15                               &  &  &  \\ \cline{1-2}
		Size of   task queue                                                                  & 10                                            &  &  &  \\ \cline{1-2}
		Size of task   input                                                                  & {[}0.2, 1{]} Mb                               &  &  &  \\ \cline{1-2}
		Speed of vehicle                                                                      & {[}30, 50{]}, {[}50, 80{]}, {[}30, 80{]} Km/h &  &  &  \\ \cline{1-2}
		RSU's   coverage range                                                                & 500 m                                         &  &  &  \\ \cline{1-2}
		RSU's   bandwidth                                                                     & 100 MHz                                       &  &  &  \\ \cline{1-2}
		V2V's   bandwidth                                                                     & 1GHz                                       &  &  &  \\ \cline{1-2}
		Channel model                                                                         & Typical Urban                                 &  &  &  \\ \cline{1-2}
		\begin{tabular}[c]{@{}c@{}}Transmission power between\\  vehicle and RSU\end{tabular} & 0.5 W                                         &  &  &  \\ \cline{1-2}
		\begin{tabular}[c]{@{}c@{}}Transmission power between vehicles \end{tabular} & 1 W                                         &  &  &  \\ \cline{1-2}
		\begin{tabular}[c]{@{}c@{}}Computation   capacity  of VEC server\end{tabular}       & [8, 14] G Cycles/s                                   &  &  &  \\ \cline{1-2}
		\begin{tabular}[c]{@{}c@{}}Computation   capacity  of vehicle\end{tabular}          & [1.8, 3.6] G Cycles/s                &  &  &  \\ \cline{1-2}
		Computation   density                                                                 & {[}20, 50{]} Cycles/bit                       &  &  &  \\ \cline{1-2}
		Waiting   time of hold on                                                             & 10, 20, 40 ms                                     &  &  &  \\ \cline{1-2}
		Delay Threshold                                 & 10, 40, 100ms                                     &  &  &  \\ \cline{1-2}		
		\begin{tabular}[c]{@{}c@{}}Output data   size/ input data size ratio\end{tabular}  & 0.05                                           &  &  &  \\ \cline{1-2}
		Energy density \cite{Dinh}                                             & $1.25\times 10^{-26}$ J/Cycle                            &  &  &  \\ \cline{1-2}
		Unit price offloading to VEC server & 0.09, 0.06, 0.03 & & & \\ \cline{1-2}
		Unit price offloading to other vehicles & 0.03, 0.06, 0.09 & & & \\ \cline{1-2}
		Parameters of utility function & $\beta_1=0.8$, $\beta_2 = 0.4$ & & & \\ \cline{1-2}
		Parameters of reward                                                                  &  \begin{tabular}[c]{@{}c@{}}${{\Gamma }_{1}}=0.8$, ${{\Gamma }_{8}}=0.9$, ${{\ell }_{1}}=-0.4$, \\ ${{\ell }_{2}}=-0.2$, ${{\ell }_{3}}=0.5$,\\ ${{\Gamma }_{2}},{{\Gamma }_{3}},{{\Gamma }_{4}},{{\Gamma }_{5}}, {{\Gamma }_{6}}, {{\Gamma }_{7}}=0.5$ \end{tabular}                                       &  &  &  \\ \cline{1-2}
	\end{tabular}\label{tab:para}
\end{table}

\begin{table*}[!h]	\centering
	\caption{The Neural Network and Training Parameters}
	\begin{tabular}{|c|c|c|c|}
		\hline
		\textbf{Parameter}                               & \textbf{Value} & \textbf{Parameter}                                & \textbf{Value}  \\ \hline
		Layers of Critic Network                         & 5              & Layer Type of Critic                              & Fully Connected \\ \hline
		Neurons of First Hidden Layer for Critic Network & 1024           & Neurons of Second Hidden Layer for Critic Network & 512             \\ \hline
		Neurons of Third Hidden Layer for Critic Network & 300            & Learning Rate of Critic                           & 0.0001          \\ \hline
		Layers of Actor Network                          & 4              & Layer Type of Actor                               & Fully Connected \\ \hline
		Neurons of First Hidden Layer for Actor Network  & 500            & Neurons of Second Hidden Layer for Actor Network  & 128             \\ \hline
		Learning Rate of Actor                           & 0.0001         & Activation Function                               & Relu            \\ \hline
		Mini-batch                                       & 128            & Buffer Size                                       & 30000           \\ \hline
	\end{tabular}\label{tab:neu}
\end{table*}

\subsection{Performance  Evaluation}
We validate the algorithm performance in terms of convergence property, task completion delay and utility level under different simulation configurations. 

\noindent\textit{(1)Convergence Property}

\begin{figure}[!h]
	\centering
	\includegraphics[width=8.8cm]{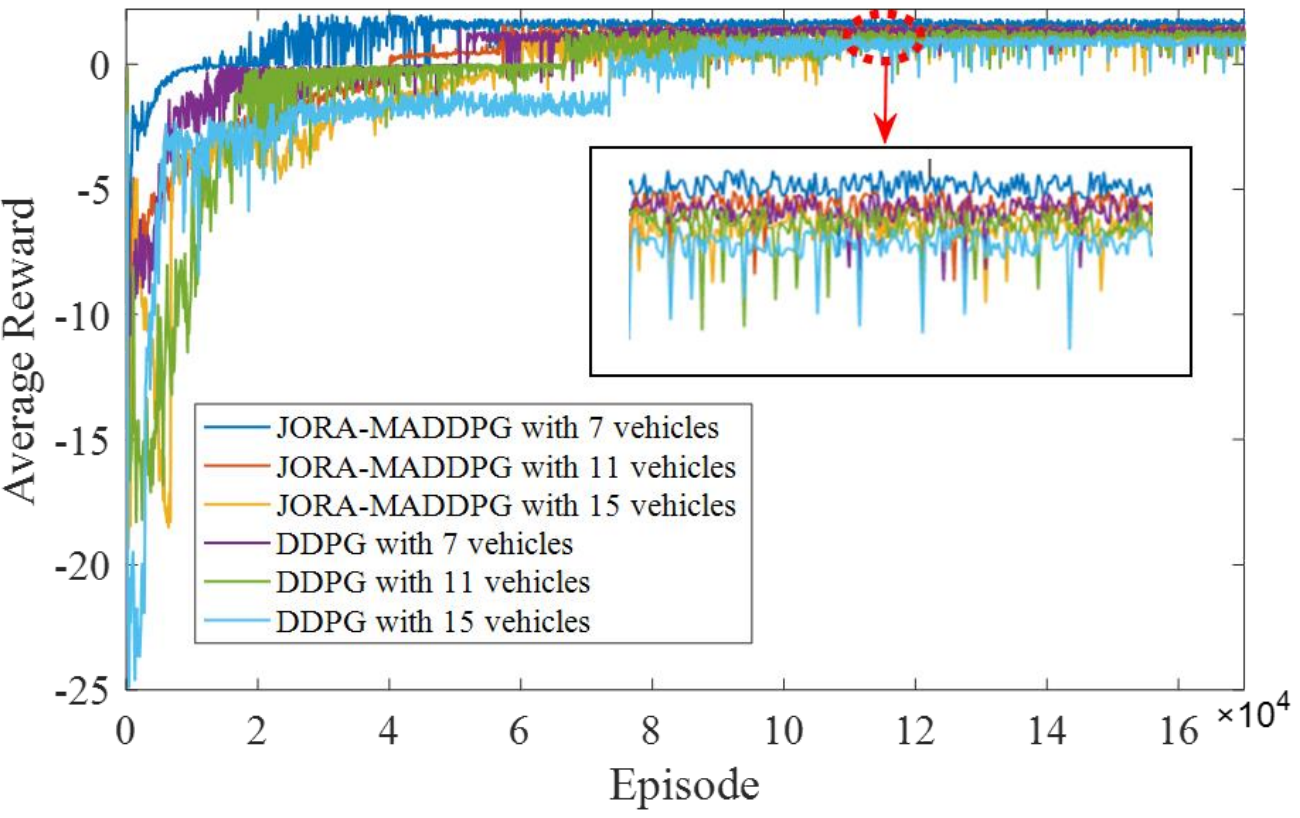}
	\caption{Convergence of different algorithms}
	\label{fig:convergence}
\end{figure}

In Fig. \ref{fig:convergence}, we present the convergence performance of our proposed JORA-MADDPG algorithm and the DDPG algorithm with the different numbers of vehicles. With the increase in the training episodes, the average reward of vehicles rises gradually and eventually preserves a stable positive reward. In the initial stage, the average reward of our proposed JORA-MADDPG algorithm and the DDPG algorithm with fewer vehicles is higher than the average reward with more vehicles. Because the increase in vehicles means a higher dimension of state space and action space, the reinforcement learning methods need to take more explorations to achieve higher rewards. As the training episodes increase, our proposed JORA-MADDPG algorithm and the DDPG algorithm gradually achieve the convergence state, where the average reward of 7 vehicles is the highest, and the average reward of 15 vehicles is the lowest because more vehicles will compete for the limited computation resources and wireless resources more fiercely, which causes the average reward to decrease. In addition, the average reward of our proposed JOAR-MADDPG algorithm is higher than the average reward of the DDPG algorithm with the same number of vehicles when the algorithms achieve the stable state because our JORA-MADDPG designs different actor-critic networks for each vehicle compared with the fact that the DDPG algorithm designs only one actor-critic network for all vehicles, which can better utilize the characteristics of the vehicle and help each vehicle achieve the near optimal offloading and resource allocation strategy.

\noindent\textit{(2)Task Completion Delay Comparison}

In Fig. \ref{fig:delay}, we present the comparison of the average task completion delay of different algorithms with different computing capacities of the VEC server. Our proposed JORA-MADDPG algorithm can guarantee that the average task completion delay of most vehicles is lower than the average task completion delay of AL, AO, RD, Greedy and DDPG algorithms because our proposed JORA-MADDPG algorithm can allocate the computation resources and wireless resources to each vehicle more accurately based on the task priority, file size, vehicle speed and vehicle channel state. In addition, the average task completion delay of vehicle 1 and vehicle 2 is significantly lower than the average task completion delay of other vehicles, and the average task completion delay of vehicle 4 is obviously higher than the average task completion delay of any other vehicle because the computing tasks of vehicle 1 and vehicle 2 are mainly CA tasks that need ultra-low delay, while the computing tasks of vehicle 4 are mainly LPA tasks whose delay requirements are not tight. 
\begin{figure}[!h]
	\centering
	\includegraphics[width=8.8cm]{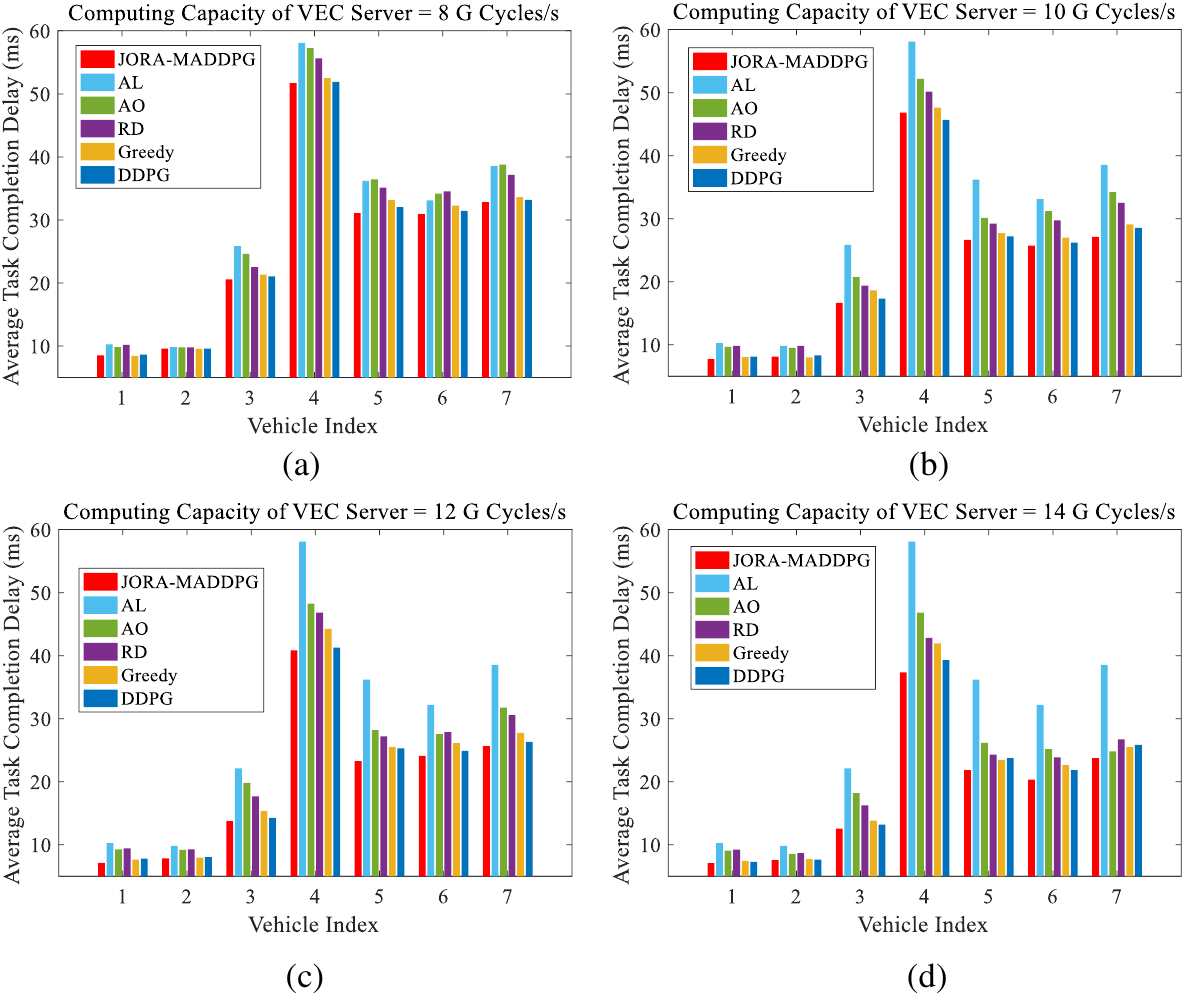}
	\caption{Average task completion delay of different algorithms: (a) the computing capacity of VEC server is 8 G Cycles/s, (b) the computing capacity of VEC server is 10 G Cycles/s, (c) the computing capacity of VEC server is 12 G Cycles/s, (d) the computing capacity of VEC server is 14 G Cycles/s.}
	\label{fig:delay}
\end{figure}

\begin{figure}[!h]
	\centering
	\includegraphics[width=7cm]{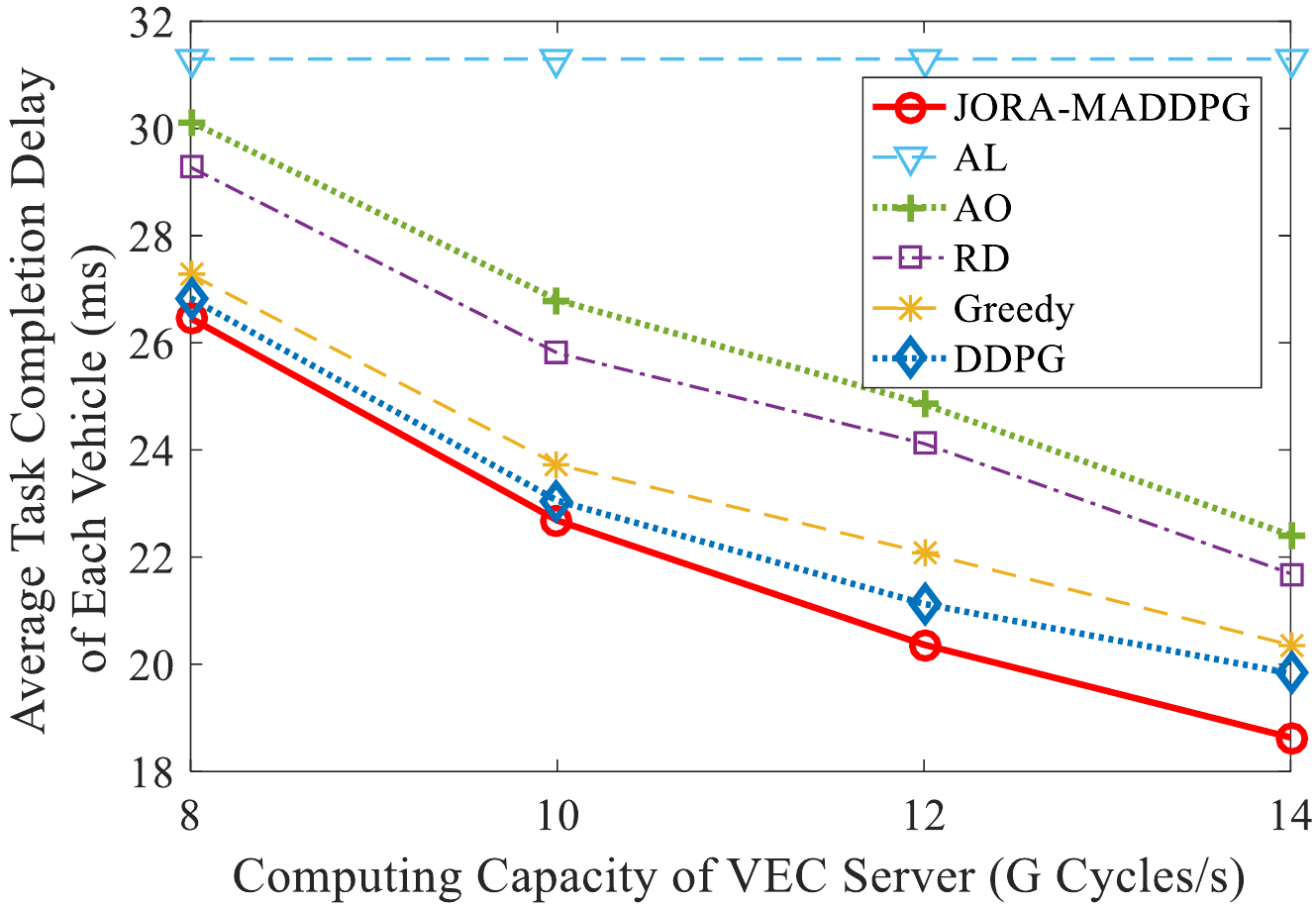}
	\caption{Average task completion delay of each vehicle vs. the different computing capacity of VEC server.}
	\label{fig:delay_avg}
\end{figure}

In Fig. \ref{fig:delay_avg}, with the increase in the computing capacity of the VEC server, the average task completion delay gradually decreases except for the AL algorithm, because computing tasks can be processed more quickly when offloaded to the VEC server. In addition, the delay gap between our proposed JORA-MADDPG algorithm and DDPG algorithm becomes wider, because our proposed algorithm can better utilize the increasing computing capacity of the VEC server to adjust the offloading and resource allocation strategy for each vehicle to further reduce the average task completion delay. 

\noindent\textit{(3)Vehicle's Utility Comparison}

\begin{figure}[!h]
	\centering
	\includegraphics[width=8.8cm]{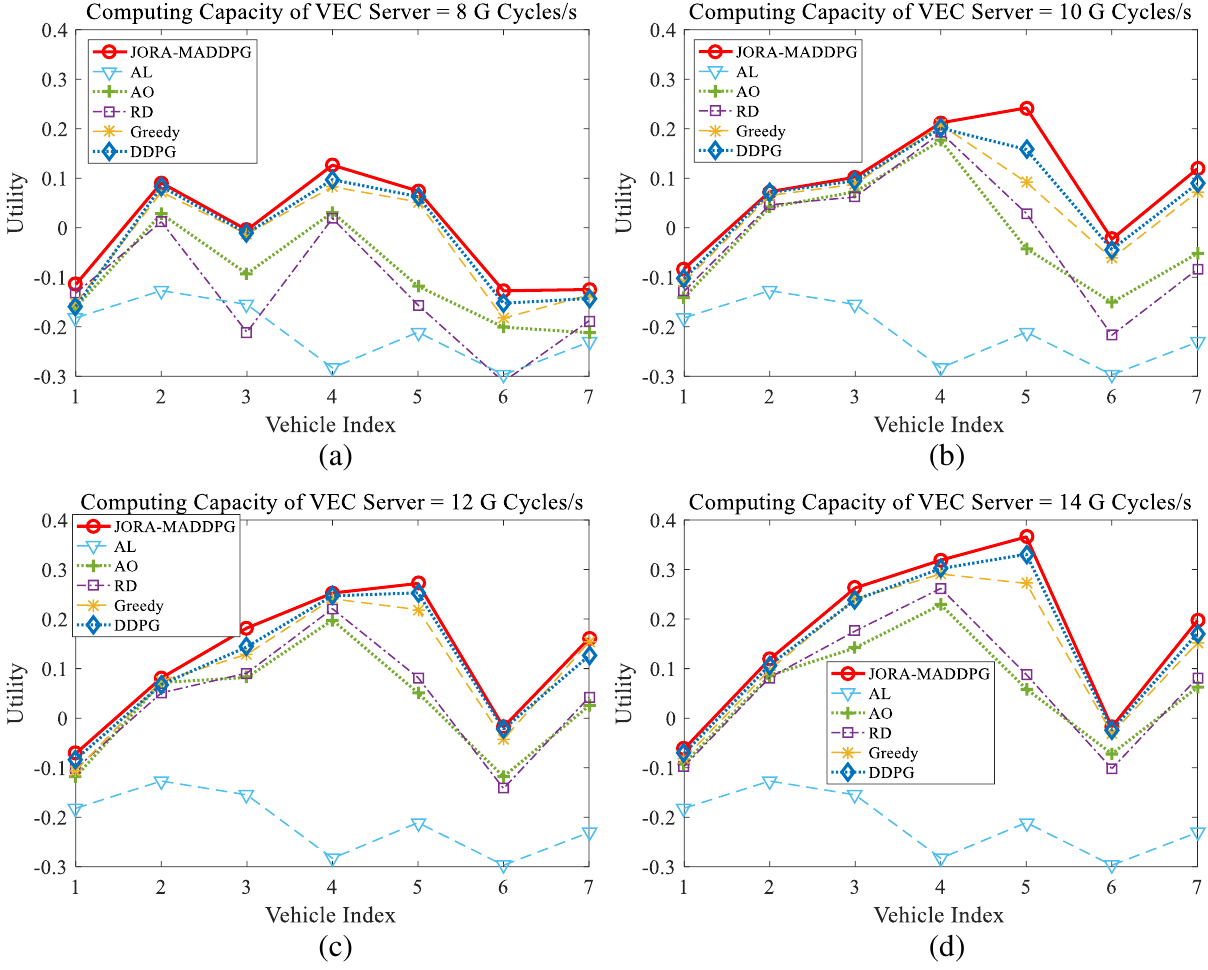}
	\caption{Utility level of different algorithms: (a) the computing capacity of VEC server is 8 G Cycles/s, (b) the computing capacity of VEC server is 10 G Cycles/s, (c) the computing capacity of VEC server is 12 G Cycles/s, (d) the computing capacity of VEC server is 14 G Cycles/s.}
	\label{fig:utility}
\end{figure}

\begin{figure}[!h]
	\centering
	\includegraphics[width=7cm]{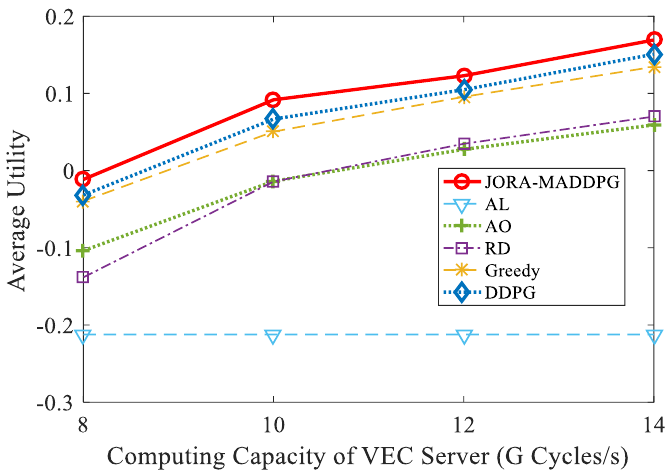}
	\caption{Average utility of each vehicle vs. the different computing capacity of VEC server.}
	\label{fig:utility_avg}
\end{figure}

In Fig. \ref{fig:utility} (a)-(d), we show the comparison of the utility level of each vehicle with different algorithms with different computing capacities of the VEC server. The comparison of average utility of all vehicles of different algorithms is also presented in Fig. \ref{fig:utility_avg}. Compared with other algorithms, our proposed JORA-MADDPG algorithm can always preserve a higher utility level for each vehicle. This is because that the vehicle's utility function is formulated into the objective function, and our reward function is designed based on the objective function and constraints, which requires the VEC server to make the near optimal offloading and resource allocation strategy for each vehicle to maximize the objective function. Therefore, the utility level of JORA-MADDPG algorithm can always be higher than the utility level of other algorithms. With the increase in the computing capacity of the VEC server, the average utility of all vehicles gradually decreases except for the AL algorithm because more computing tasks are offloaded to the VEC server to decrease the local processing energy consumption.

\noindent\textit{(4)Analysis of the Number of Vehicles on Task Completion Delay and Vehicle's Utility}

\begin{figure}[!h]
	\centering
	\includegraphics[width=8.8cm]{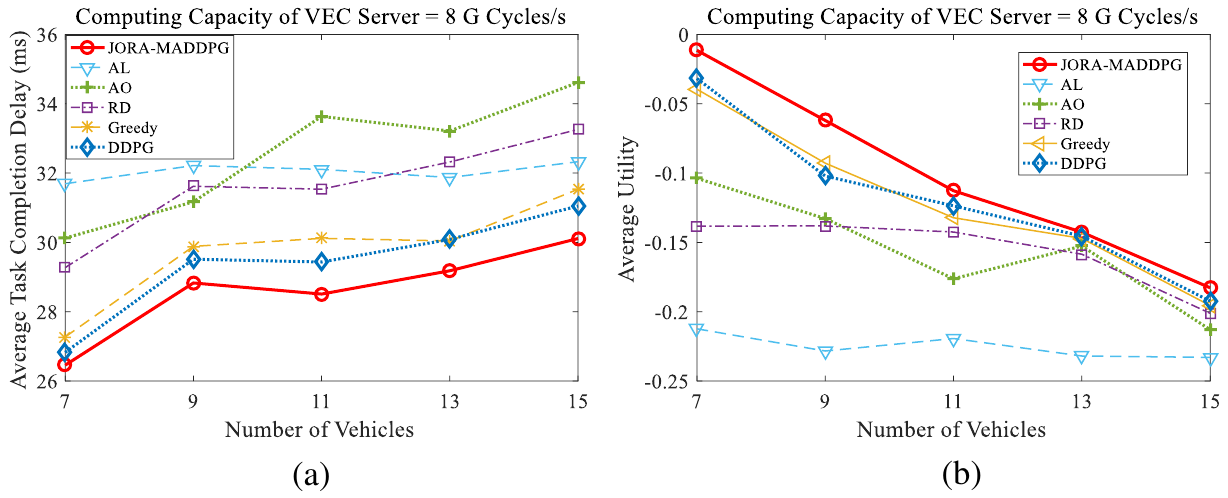}
	\caption{Average task completion delay and average utility vs. the number of vehicles: (a) the average task completion delay when the computing capacity of VEC server is 8 G Cycles/s, (b) the average utility when the computing capacity of VEC server is 8 G Cycles/s.}
	\label{fig:com}
\end{figure}

In Fig. \ref{fig:com} (a) and (b), we present the average task completion delay and average utility under the different number of vehicles, respectively. With the increase in the number of vehicles, the average task completion delay of all algorithms rises, and the average utility of all algorithms declines except for the AL algorithm because the increasing vehicles will generate more computing tasks to compete for drastically limited computation resources and wireless resources from the VEC server, which indicates that some computing tasks cannot be processed in a timely manner. Therefore, the average task completion delay gradually rises. However, more computing tasks are selected to execute on the local vehicle terminal and other vehicle terminals because of the limited computation resources and wireless resources of the VEC server. However, the cost of energy consumption of the vehicle terminals exceeds the cost of the revenue from the processing tasks. Therefore, the average utility of the vehicles gradually decreases. 

\noindent\textit{(5)Analysis of the Vehicle Speed on Task Completion Delay and Vehicle's Utility}

\begin{figure}[!h]
	\centering
	\includegraphics[width=8.8cm]{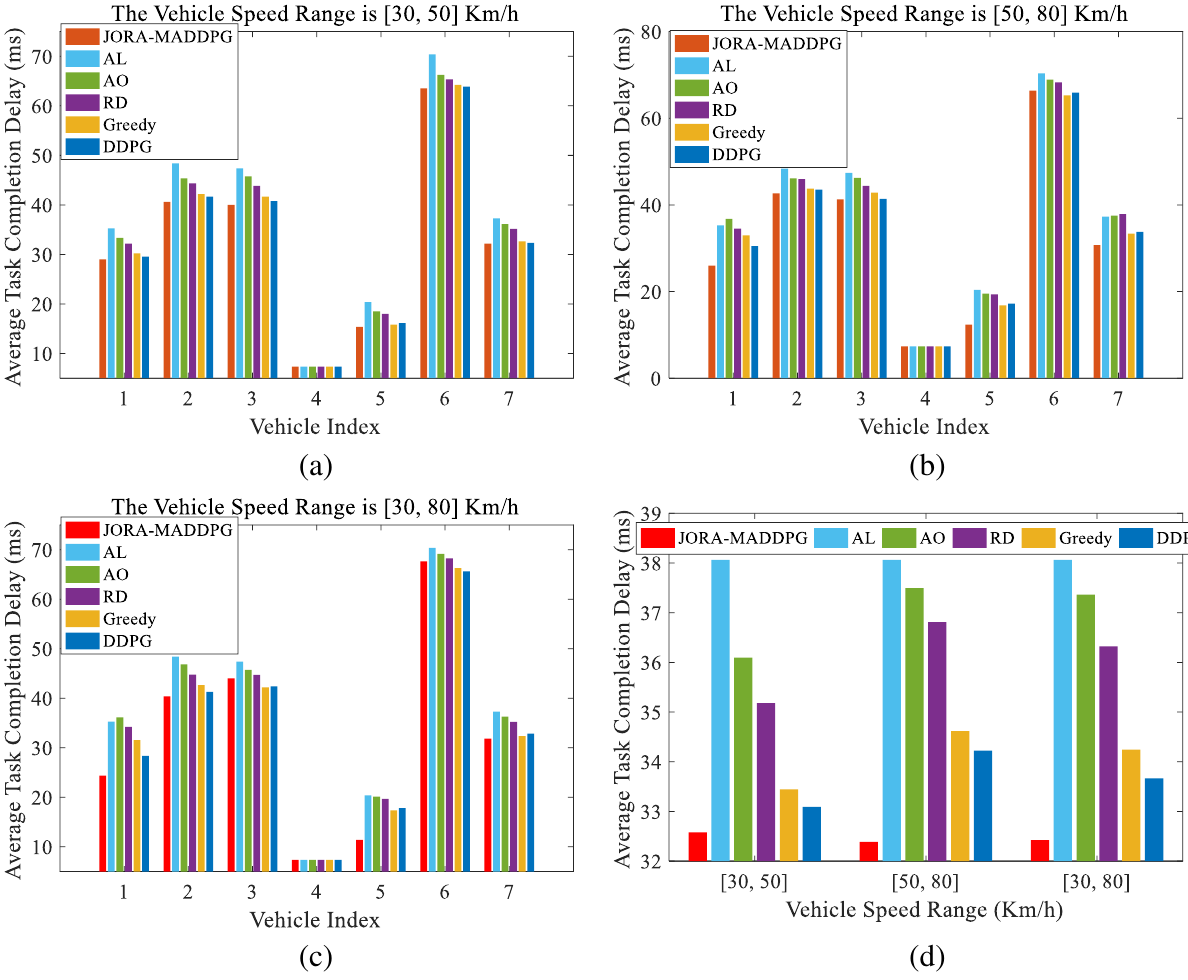}
	\caption{Average task completion delay of different algorithms under different vehicle speed ranges: (a) the vehicle speed range is [30, 50] Km/h, (b) the vehicle speed range is [50, 80] Km/h, (c) the vehicle speed range is [30, 80] Km/h, (d) average task completion delay of different vehicle speed ranges.}
	\label{fig:12}
\end{figure}

In Fig. \ref{fig:12}, we compare the average task completion delay under different vehicle speed ranges. Compared with AL, AO, RD, Greedy and DDPG algorithms, our proposed JORA-MADDPG algorithm can significantly reduce the average task completion delay of most vehicles because our optimization goal is to maximize the utility level of the vehicles within task delay constraints. As long as the task completion delay can satisfy the delay requirement, the computing task can be finished to satisfy the requirements of drivers and passengers. In addition, the average task completion delay of vehicle 4 is significantly lower than the average task completion delay of other vehicles, and the average task completion delay of vehicle 6 is obviously higher than the average task completion delay of any other vehicle because the computing tasks of vehicle 4 are mainly CA tasks that need ultra-low delay, while the computing tasks of vehicle 6 are mainly LPA tasks whose delay requirements are not tight. Through the average task completion delay comparison of different vehicle speed ranges, the average task completion delay of the vehicle speed range [30, 80] is the lowest because the large vehicle speed difference leads to a large difference in the task delay constraint, which can guide the VEC server to allocate more wireless and computation resources to the computing task with lower delay constraint.

\begin{figure}[!h]
	\centering
	\includegraphics[width=8.8cm]{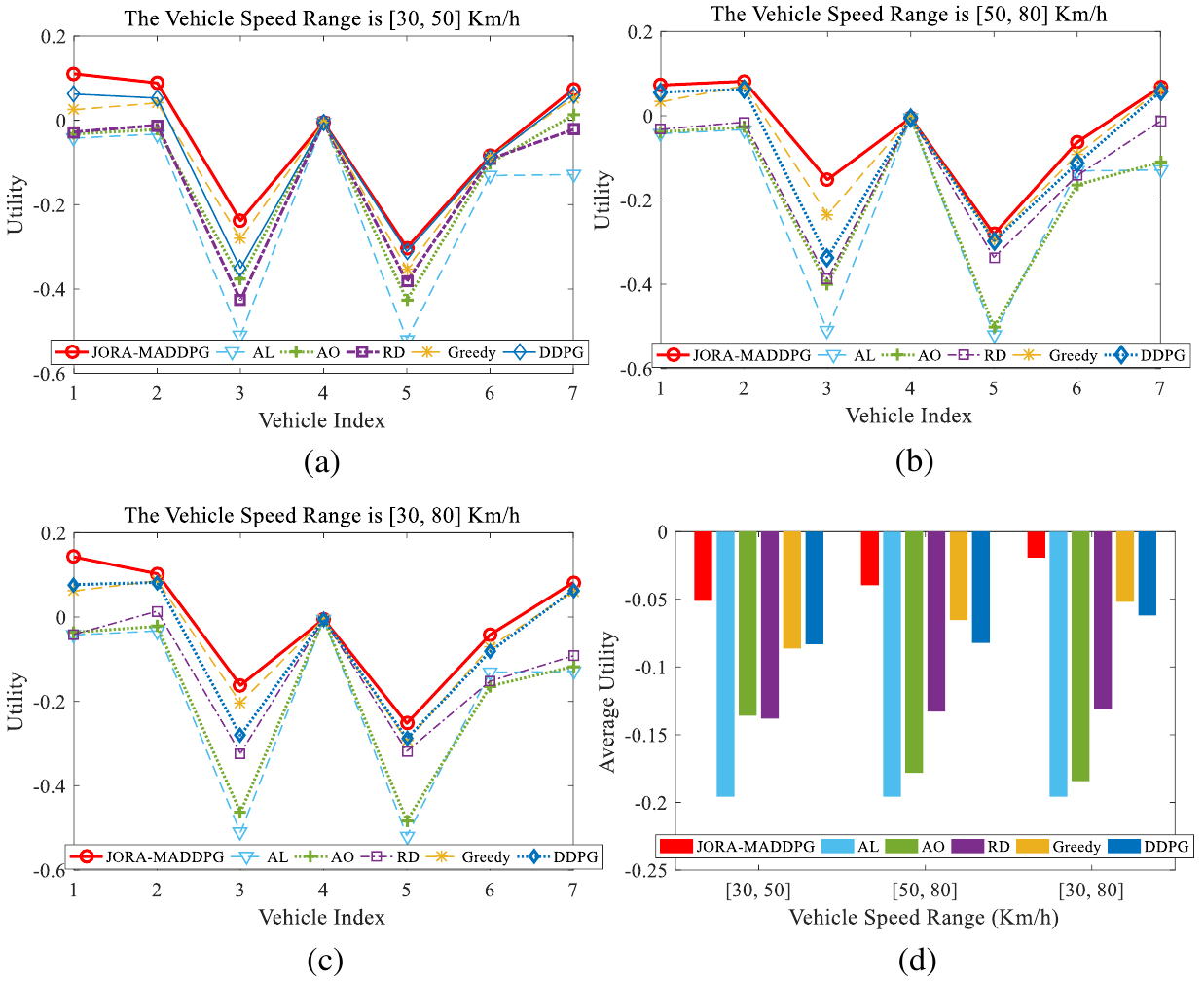}
	\caption{Utility of different algorithms under different vehicle speed ranges: (a) the vehicle speed range is [30, 50] Km/h, (b) the vehicle speed range is [50, 80] Km/h, (c) the vehicle speed range is [30, 80] Km/h, (d) average utility of different vehicle speed ranges. The computing capacity of VEC server is 8 G Cycles/s}
	\label{fig:13}
\end{figure}

In Fig. \ref{fig:13}, we compare the vehicle's utility of the vehicle and the average utility of all the vehicles under at different vehicle speed ranges. Compared with the AL, AO, RD, Greedy and DDPG algorithms, our proposed JORA-MADDPG algorithm can significantly increase the vehicle's utility of the vehicle. This is because that our utility function is formulated by the energy consumption and revenue of the processing tasks. By offloading the vehicles' computing tasks of the vehicles to the VEC sever, the vehicle's energy consumption of the vehicle for processing tasks can be maintained at a low level. In addition, based on the collaborated offloading mechanism among vehicles, the vehicle's computing tasks of the vehicle can also be offloaded to the a vehicle with available computation resources, so the vehicle can benefit from processing tasks for other vehicles. Through the average utility comparison of different vehicle speed ranges, it can be observed that the average utility of vehicle speed range [30, 80] is seen to be the highest. The reason is that because the vehicles of at a high speed can offload their computing tasks to the a vehicle of low speed to process through the V2V communication, and the task delay constraint of the vehicle of low speed is relatively loose. Therefore, the offloading behaviors between vehicles becomes more frequently, and the processing revenue benefits a lot. In this case, the average utility is relatively high.

\noindent\textit{(6)Analysis of the Offloading Unit Price on Task Completion Delay and Vehicle's Utility}

\begin{figure}[!h]
	\centering
	\includegraphics[width=8.8cm]{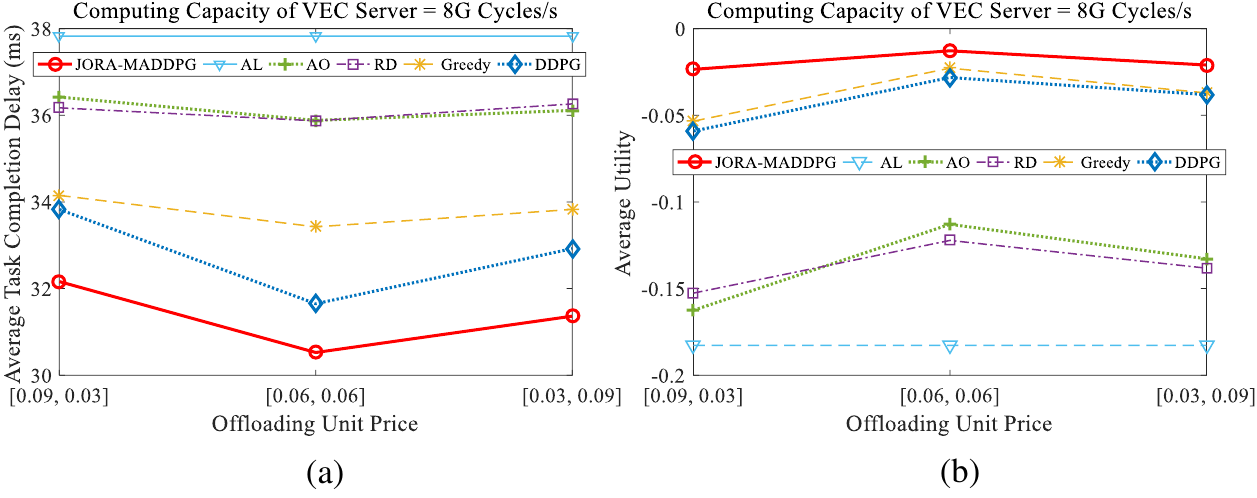}
	\caption{Average task completion delay and average utility of different algorithms under different offloading unit prices: (a) average task completion delay of different offloading unit prices, (b) average utility of different offloading unit prices.}
	\label{fig:14}
\end{figure}

In Fig. \ref{fig:14}, we compare the average task completion delay and average utility of all vehicles under different offloading unit prices. For the offloading unit price, the unit price of offloading to the VEC server and other vehicles is expressed in the same set. For instance, [0.09, 0.03] denotes that the unit price of offloading computing tasks to the VEC server and other vehicles is 0.09 and 0.03, respectively. Compared with other algorithms, our proposed JORA-MADDPG algorithm can always preserve a lower average task completion and a higher average utility under different offloading unit prices because our proposed algorithm can adjust the offloading and resource allocation strategy for each vehicle when the offloading unit price changes. Specifically, when the unit price of offloading tasks to the VEC server is higher, the computing tasks of the vehicle are preferably processed locally or on the terminals of other vehicles through V2V communication. In this case, the average completion delay is relatively high compared with offloading tasks to the VEC server, but the average utility of the vehicles is higher because of the increased processing revenue. With a decrease in the unit price of offloading tasks to the VEC server, the computing tasks of the vehicles are preferentially offloaded to the VEC server to process for the lower task completion delay and less energy consumption.

\section{Conclusion}
In this paper, we propose a joint task type and vehicle speed-aware computing task offloading and resource allocation algorithm to achieve the goal of energy-efficiency and increasing revenues for all vehicles within task delay constraints. First, the task delay constraint model is established based on task type and vehicle speed. Then, we calculate the task completion delay, energy consumption and processing revenue for different offloading patterns based on the allocated computation and wireless resources. Based on the energy consumption and processing revenue, the utility function of the vehicle is formulated. Finally, the joint optimization of the offloading and resource allocation problem is formulated by the Markov decision process (MDP), with the objective of maximizing the utility of the vehicles subject to the delay constraints. Based on the objective function and constraints, we proposed the JORA-MADDPG algorithm to obtain a near-optimal solution to this problem. Simulation results show that our proposed JORA-MADDPG algorithm can effectively decrease energy consumption and task completion delay and improve the revenue of the vehicles compared with other algorithms.

\section*{Acknowledgment}
This research work was supported in part by the National Science Foundation of China (U1903213, 61701389) and the Shaanxi Key R\&D Program (2018ZDCXL-GY-04-03-02).

\ifCLASSOPTIONcaptionsoff
  \newpage
\fi

\end{document}